\documentclass[twocolumn]{aastex631}

\newcommand{\kms}        {\ifmmode{\rm \,km\,s^{-1}}\else\,km\,s$^{-1}$\xspace\fi}
\shorttitle{Raman-scattered He~II in Planetary Nebulae}
\shortauthors{Lim et al.}
\graphicspath{{./}{figures/}}
\usepackage{multirow}
\usepackage{array}
\usepackage{subfigure}
\usepackage{threeparttable}
\usepackage{appendix}
\usepackage{comment}
\begin{document}

\title{ High Resolution {\it BOES} Spectroscopy of Raman-scattered He~II$\lambda$6545 
in Young Planetary Nebulae}

\author[0009-0007-6875-481X]{Jin Lim}
\affiliation{Department of Physics and Astronomy, Sejong University, Seoul, Korea}

\author[0000-0002-0112-5900]{Seok-Jun Chang}
\affiliation{Max-Planck-Institut f\"{u}r Astrophysik, Karl-Schwarzschild-Stra$\beta$e 1, 85748 Garching b. M\"{u}nchen, Germany}

\author[0000-0001-6363-8069]{Jaejin Shin}
\affiliation{Department of Physics and Astronomy, Sejong University, Seoul, Korea}

\author[0000-0002-1951-7953]{Hee-Won Lee}
\affiliation{Department of Physics and Astronomy, Sejong University, Seoul, Korea}

\author{Jiyu Kim}
\affiliation{Department of Physics and Astronomy, Sejong University, Seoul, Korea}

\author[0000-0001-7033-4522]{Hak-Sub Kim}
\affiliation{Department of Physics and Astronomy, Sejong University, Seoul, Korea}
\affiliation{Korea AeroSpace Administration, Sacheon, Korea}

\author[0000-0002-9040-672X]{Bo-Eun Choi}
\affiliation{Department of Astronomy, University of Washington, Seattle, WA 98195, USA}

\author[0000-0002-3808-7143]{Ho-Gyu Lee}
\affiliation{Korea Astronomy and Space Science Institute, Daejeon, Korea}

\begin{abstract}

Young planetary nebulae (PNe) are characterized by their hot central stars and the presence of abundant neutral and molecular components, which result from significant mass loss during the asymptotic giant branch (AGB) phase of stellar evolution. Far-UV \ion{He}{2}$\lambda$1025 line photons produced near the central star can undergo Raman scattering by hydrogen atoms, creating a broad emission feature centered at $\sim$ 6545~\AA.
We conducted high-resolution spectroscopy of 12 young PNe from April 2019 to March 2020 using the Bohyunsan Observatory Echelle Spectrograph ({\it BOES}). Building on the study by Choi and Lee, who identified Raman-scattered \ion{He}{2} at 6545~\AA\ in NGC~6881 and NGC~6886, we report new detections of this feature in NGC~6741 and NGC~6884.
Profile fitting reveals that the velocity of the \ion{H}{1} component relative to the \ion{He}{2} emission region ranges from $26-33~{\rm km~s^{-1}}$ in these PNe. Using photoionization modeling, we estimate the line flux of \ion{He}{2}$\lambda$1025 and derive Raman conversion efficiencies of 0.39, 0.21, 0.24, and 0.07 for NGC~6881, NGC~6741, NGC~6886, and NGC~6884, respectively. These results, combined with radiative transfer modeling, suggest the presence of \ion{H}{1} components with masses around $10^{-2}~M_\odot$, moving outward from the central \ion{He}{2} emission region at speeds characteristic of the slow stellar wind from a mass-losing giant star.
\end{abstract}

\keywords{ Interdisciplinary astronomy(804)}

\section{Introduction} \label{sec:intro}

A star with mass $\le 8\ {M_\odot}$ loses a significant amount of its mass leaving behind a hot
central core to become a planetary nebula (PN). A PN is characterized by an emission nebula that is photoionized by
the hot central star. 
Young PNe are particularly important in the study of the mass loss process that characterizes
the stellar evolution in the AGB stage. 
The slow stellar wind in the AGB stage mainly consists
of molecules and dust grains with a significant contribution of neutral atomic species.

The strong far-UV radiation from the hot central star is believed to dissociate a significant amount of molecules into neutral species \citep[e.g.,][]{schneider87}. \citet{natta98} conducted extensive theoretical investigations into the evolution of neutral and molecular components in PNe. According to their modeling, the slowly expanding shell, primarily composed of dust and molecules, is further driven by the hot tenuous wind to reach a final speed of $\sim 25~{\rm km~s^{-1}}$. 
As photodissociation progresses, the \ion{H}{1} mass gradually increases, eventually reaching a maximum value of a few times $10^{-2}~M_\odot$ in the case of a central star with mass $M_*=0.6~M_\odot$. 
However, 21 cm radio observations of \ion{H}{1} are severely hindered due to significant confusion from interstellar emission unless the object exhibits a substantial peculiar velocity relative to Galactic rotation \citep[e.g.,][]{schneider87, taylor89, taylor90}. This limitation makes it challenging to trace neutral hydrogen components in a direct and reliable way \citep[e.g.,][]{hofner18}.

Young PNe are also characterized by their compact size and the presence of high-excitation lines, including \ion{He}{2}, which are produced by the central hot star. Considering the ionization potential of \ion{He}{2} exceeds 50~eV, the \ion{He}{2} emission nebula is photoionized by an extremely hot source. As a PN evolves, the nebula expands, and the temperature of the hot source decreases, making the duration of \ion{He}{2} emission relatively short compared to the entire lifespan of the PN stage.
Strong \ion{He}{2} emission and a compact size create favorable conditions for the Raman scattering of \ion{He}{2} with atomic hydrogen to occur.

The first detection of Raman-scattered \ion{He}{2} features in PNe was made by \citet{pequignot97} in the young PN NGC~7027. Subsequently, Raman-scattered \ion{He}{2} features were identified in three young PNe: NGC~6302, IC~5117, and NGC~6790 \citep{groves02, lee06, kang09}. 
The detection of these spectral features indicates the presence of thick \ion{H}{1} components with $N_{\rm HI}\sim 10^{21-22}~{\rm cm^{-2}}$. Raman-scattered \ion{He}{2} features at 4850~\AA\ in IC~5117 and NGC~6790 suggest that the \ion{H}{1} components are moving away from the \ion{He}{2} emission region with expansion speeds of $v_{\rm exp}= 30$ and $20~{\rm km~s^{-1}}$, respectively. From theoretical modeling of Raman-scattered \ion{He}{2} formation using a Monte Carlo technique, the Raman conversion efficiency of \ion{He}{2}$\lambda$6545, amounting to $\sim 10\%$, is consistent with an \ion{H}{1} mass of $\sim 10^{-2}~M_\odot$ \citep[e.g.,][]{choi20a}.

Recently, \citet{choi20b} reported the discovery of Raman-scattered \ion{He}{2} at 6545~\AA\ in NGC~6881 and NGC~6886. In this article, we present high-resolution spectroscopy of 12 young PNe and report our discovery of Raman-scattered \ion{He}{2} at 6545~\AA\ in two additional objects, NGC~6741 and NGC~6884, along with NGC~6881 and NGC~6886.
In Section~2, we describe the basic atomic physics underlying the formation of Raman-scattered \ion{He}{2}$\lambda$6545 and its spectroscopic identification. In Sections~3 and 4, we present our spectra and perform line fit analysis to derive the relative velocity between the \ion{He}{2} emission region and the \ion{H}{1} region. In Section~5, we conduct photoionization computations and radiative transfer modeling to deduce the Raman conversion efficiency and estimate the \ion{H}{1} mass.
In the following section, we discuss the importance of Raman spectroscopy as a unique probe of PN evolution. Finally, in the last section, we summarize our results and outline directions for future work.

\section{Formation of Raman-scattered He~II}

% \subsection{Atomic Physics }

Raman scattering involving atomic hydrogen offers a novel window for studying the distribution and kinematics of \ion{H}{1} in young PNe. Raman scattering of a far-UV photon blueward of Ly$\alpha$ occurs when the photon interacts with a hydrogen atom in the ground state. This interaction causes the atom to de-excite to the $2s$ state while emitting a photon whose energy is reduced by the Ly$\alpha$ transition energy. As a single-electron species, \ion{He}{2} has an energy level structure resembling that of atomic hydrogen. Nevertheless, the spacing between its energy levels is approximately four times wider. \ion{He}{2} spectral lines from $2n\to2$ transitions have shorter wavelengths than the corresponding \ion{H}{1} Lyman series ($n\to 1$). This difference arises because the reduced mass of \ion{He}{2} is larger than that of \ion{H}{1} by approximately the electron-to-proton mass ratio.

\begin{figure}
    \centering
    \includegraphics[scale=0.3]{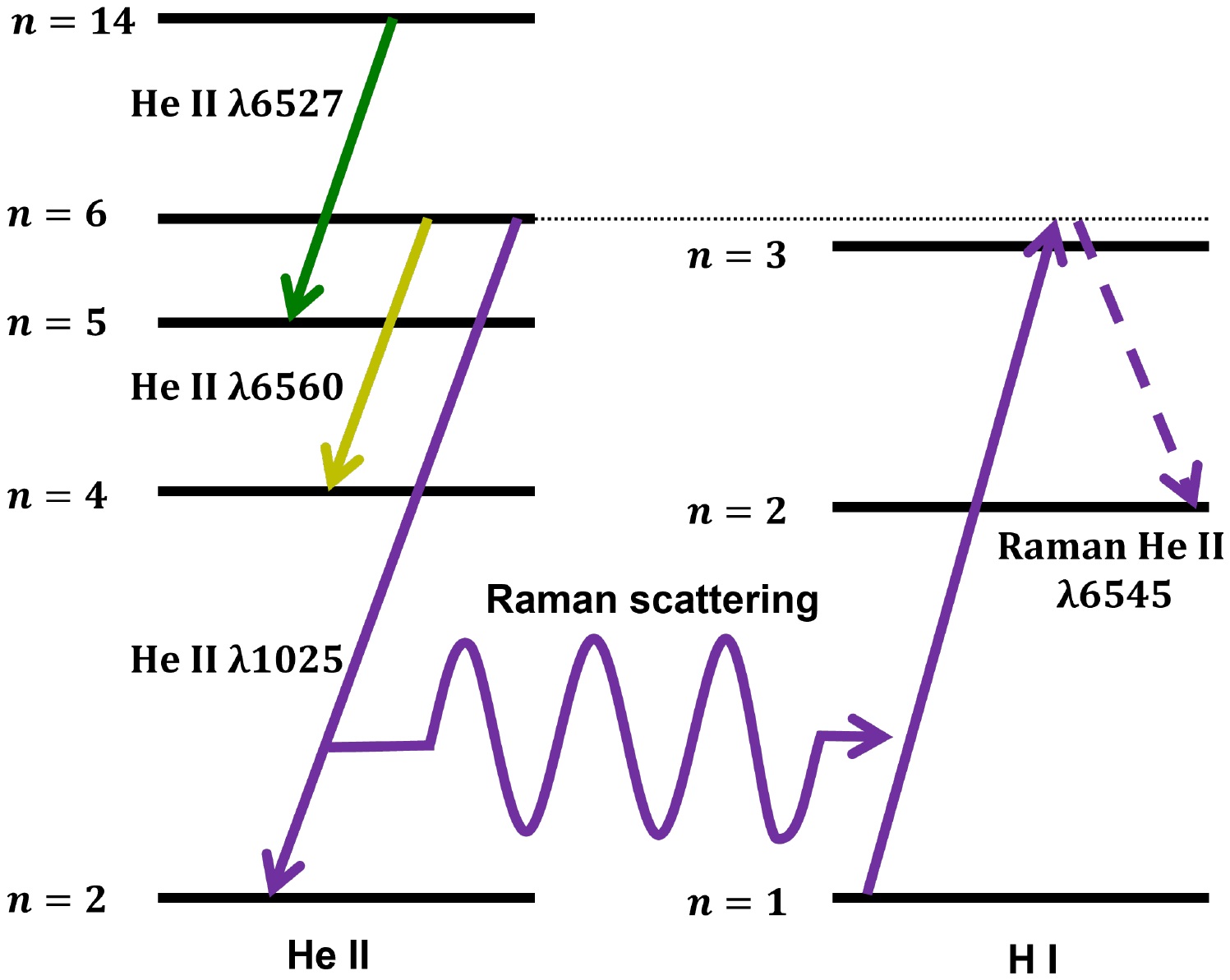}
    \caption{The energy levels of \ion{He}{2} and \ion{H}{1} relevant to formation of Raman-scattered \ion{He}{2} at 6545\AA. The green and yellow arrows represent optical \ion{He}{2} emission lines near Raman-scattered \ion{He}{2}.}
    \label{fig:ram6545}
\end{figure}

Quantitatively, the energy difference for \ion{H}{1} and \ion{He}{2} can be expressed as
\begin{equation}
\Delta E_n \simeq \frac{3}{4} \left(\frac{m_e}{m_p}\right) E_{\rm ryd} (1-n^{-2}),
\end{equation}
where $E_{\rm ryd}$ denotes the Rydberg energy for hydrogen, and $m_e$ and $m_p$ represent the masses of the electron and proton, respectively.
Thus, Raman scattering of \ion{He}{2} produces an optical emission feature blueward of the nearby hydrogen Balmer line, with the wavelength difference given by
\begin{equation}
\Delta \lambda \simeq -5.9 \left[ \frac{n^2 (n^2 -1)}{(n^2 - 4)^2} \right]\, {\rm \AA}.
\end{equation}
\cite{lee01}.
Specifically, for Raman scattering of \ion{He}{2}\,$\lambda 1025$, originating from the $6 \to 2$ transition, we find $\Delta \lambda = -18\, {\rm \AA}$ with $n=3$, leading to the formation of an optical line feature at 6545~\AA.
Figure~\ref{fig:ram6545} presents an energy level diagram illustrating the formation of Raman-scattered \ion{He}{2} at 6545~\AA.
Interestingly, a spectral line, \ion{He}{2}\,$\lambda 6527$, corresponding to the $n=14 \to 5$ transition, is observed with a strength comparable to that of Raman-scattered \ion{He}{2} at 6545~\AA\ in young PNe \citep[e.g.,][]{lee00, choi20b}.

A very important spectroscopic feature of Raman scattering is that the Raman line profile width can differ significantly from that of the incident radiation. 
This difference arises from the transformation of wavelength spaces, as incident radiation in the far-UV regime is shifted to the optical region as Raman-scattered radiation. 
The energy conservation principle in the Raman scattering process establishes the relation:
\begin{equation}
  \nu_{\rm inc} = \nu_{\rm Ram} + \nu_{\rm Ly\alpha},
\label{eq:Ram_Wave}
\end{equation}
where $\nu_{\rm inc}$, $\nu_{\rm Ram}$, and $\nu_{\rm Ly\alpha}$ are the frequencies of the incident, Raman-scattered, and Ly$\alpha$ photons, respectively. Since $\nu_{\rm Ly\alpha}$ is fixed, the variations in $\nu_{\rm inc}$ and $\nu_{\rm Ram}$ are equal. Consequently, the line widths of \ion{He}{2}$\lambda$1025 and Raman-scattered \ion{He}{2}$\lambda$6545 are related as:
\begin{equation}
    {\Delta\nu_{\rm inc} \over \nu_{\rm inc}} = \left( {\nu_{\rm Ram} \over \nu_{\rm inc}} \right) 
    {\Delta\nu_{\rm Ram} \over \nu_{\rm Ram}} = {1025 \over 6545} 
    {\Delta\nu_{\rm Ram} \over \nu_{\rm Ram}}.
\label{eq:Ram_Con}
\end{equation}
This relationship indicates that the line width of Raman-scattered \ion{He}{2}$\lambda$6545 is broadened by a factor of 6.4 compared to that of the \ion{He}{2} emission lines.

In addition to profile broadening, the expansion of wavelength space in Raman scattering should also be considered when determining the wavelength shift caused by the Doppler effect. When the \ion{H}{1} region moves away from the \ion{He}{2} emission region, the Raman \ion{He}{2} feature shifts redward of the atomic line center. This shift follows the same expansion effect in wavelength space as described in Equation~(\ref{eq:Ram_Con}).

For example, if the relative velocity between the \ion{H}{1} and \ion{He}{2} regions is $+10 {\rm\ km\ s^{-1}}$, the Raman \ion{He}{2}$\lambda$6545 line shifts by approximately $+1.4~$\AA\ in wavelength space. 
Given the 3\,\AA\ separation between Raman-scattered \ion{He}{2}$\lambda6545$ and [\ion{N}{2}]$\lambda6548$, a recession velocity of approximately $20 {\rm\ km\ s^{-1}}$ for the \ion{H}{1} region would result in complete blending of the two spectral features.
Since the Raman-scattered \ion{He}{2}$\lambda$6545 feature is relatively weak, it is naturally expected to appear as broad wings around the [\ion{N}{2}]$\lambda$6548 line.

\begin{figure*}
    \centering
    \includegraphics[width=\textwidth]{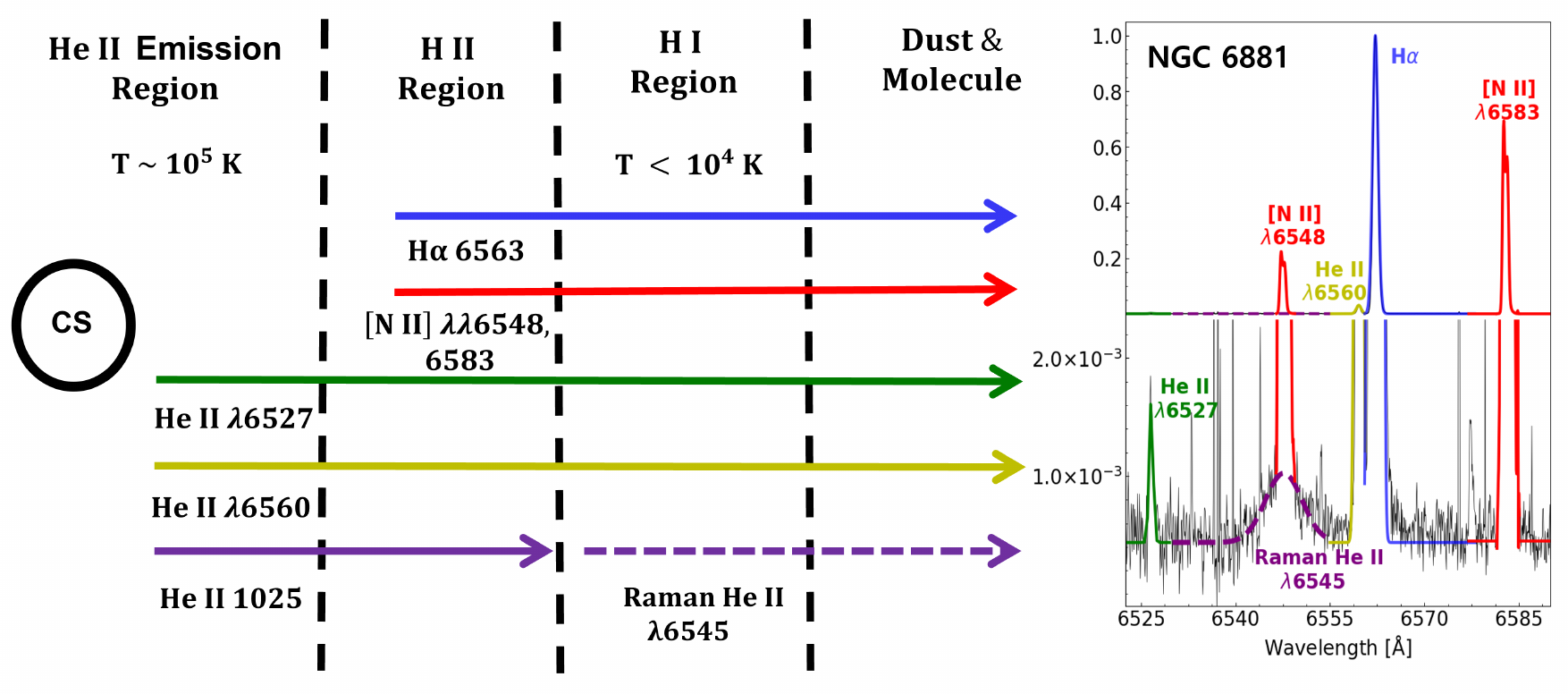}
    \caption{A schematic illustration of the formation of emission lines near H$\alpha$ (left) and the spectrum of NGC~6881 obtained  with {\it BOES} (right). The central star (denoted by 'CS' inside a circle) is sufficiently hot to create a \ion{He}{2} emission region with $T\sim 10^5 \, \rm K$. Various emission lines are shown by colored arrows: \ion{He}{2} $ \lambda6527$ (green), \ion{He}{2}~ $\lambda6560$ (yellow), [\ion{N}{2}] $\lambda\lambda6548$ and 6583 (red), $\rm H\alpha$ (blue), \ion{He}{2} $\lambda1025$, and Raman \ion{He}{2} $\lambda6545$ (purple). On the right side, the spectrum is normalized to the peak of $\rm H\alpha$. In the lower panel, the vertical scale is reduced to clearly reveal weak emission lines including \ion{He}{2} $ \lambda6527$ and Raman-scattered \ion{He}{2} $\lambda6545$, the broad wing feature (purple dashed line) around [\ion{N}{2}]$\lambda$6548. 
    }
    \label{fig: schematic_lines}
\end{figure*}

A careful spectroscopic analysis is required to detect the broad wings surrounding the [\ion{N}{2}]$\lambda$6548 emission line. Figure~\ref{fig: schematic_lines} presents a schematic illustration of the formation of various emission lines near H$\alpha$. The right panel displays {\it BOES} spectroscopic data of NGC~6881, showing broad spectral wings near [\ion{N}{2}]$\lambda$6548.
Interestingly, [\ion{N}{2}]$\lambda$6548 is accompanied by [\ion{N}{2}]$\lambda$6583, which is nearly three times stronger \citep[e.g.,][]{osterbrock89}. If [\ion{N}{2}]$\lambda$6583 does not display prominent broad wings, it strongly indicates that the wings around [\ion{N}{2}]$\lambda$6548 are not associated with nitrogen ions. This contributes to the identification of the broad wings as Raman-scattered \ion{He}{2}$\lambda$6545.

\section{Spectroscopic observation}\label{sec:spectro}

With a view to searching for Raman-scattered \ion{He}{2} features, we selected 
12 PNe from the two catalogs provided by \citet{tylenda94} and \citet{sahai11}. \citet{tylenda94} compiled the intensities of \ion{He}{2} $\lambda$4686 and H$\alpha$ lines 
for Galactic PNe, while \citet{sahai11} introduced a new morphological classification for 119 young PNe and determined their ages. 
We preferentially selected strong \ion{He}{2} emitters with line intensity ratios of \ion{He}{2} $\lambda$4686 to H$\alpha$ greater than 0.05. 
Additionally, since young PNe are believed to contain a significant amount of neutral material including atomic hydrogen, we included some young PNe even if they have relatively low-intensity ratios.

\begin{table*}[h]
    \centering
    \renewcommand{\arraystretch}{1.1}
    \caption{Observational log for the 12 young PNe and their physical parameters.} % in the literature.}
    \begin{tabular}{ccccc|c}
    \hline
    Object& Date & Total Exposure Time & $\log{T_{\rm eff}}$  &$\log{L/L_\odot}$ & Group$^{\dagger}$ \\
           &(yyyy-mm-dd) & (sec) & (K) & & \\
    \hline
    H 4-1     &    2019-04-06 & 1800  & $5.11$\textsuperscript{a} & $2.28\textsuperscript{a}$  &\multirow{3}{*}{N}  \\
    Hu 2-1    &   2020-03-30 & 2400  & $4.51$\textsuperscript{b} & $4.06\textsuperscript{b}$ &  \\
    Hen 2-447 &    2019-06-05 &  600  &  $5.03$\textsuperscript{c} & $2.77\textsuperscript{c}$  & \\
    \hline
    M 1-8     &    2020-03-30 & 3600 & $5.21$\textsuperscript{c} & $2.12\textsuperscript{c}$ & \multirow{5}{*}{H} \\
    NGC 2346  &  2019-04-05 & 1800 &  $4.76$\textsuperscript{c} & $3.84\textsuperscript{c}$  \\
    NGC 2392  &  2020-03-30 & 1200 & $4.90$\textsuperscript{d} & $4.41\textsuperscript{d}$  \\
    J 900     &  2019-04-06 & 1800 & $5.11$\textsuperscript{e} & $3.75\textsuperscript{e}$ &  \\
    NGC 3242  &    2020-03-30 & 3600   & $4.90$\textsuperscript{d} & $2.87\textsuperscript{d}$  & \\
    \hline
    NGC 6881  &    2020-03-30 & 3300  & $4.99$\textsuperscript{c} & $2.57\textsuperscript{c}$ & \multirow{4}{*}{R} \\
    NGC 6886  &    2019-10-30 & 2400  & $5.18$\textsuperscript{d} & $2.83\textsuperscript{d}$    &  \\
    NGC 6741  &   2020-03-28 & 3600  & $5.23$\textsuperscript{f}  & $2.75\textsuperscript{f}$   &  \\  
    NGC 6884  &    2020-03-28 & 3600 & $4.90$\textsuperscript{c} & $3.27\textsuperscript{c}$     & \\
    \hline
    \end{tabular}
    \parbox{0.6\textwidth}{
    \footnotesize
    $\dagger$: Three groups as shown in Figure~\ref{fig:represent}.
    \textbf{a:} \cite{otsuka23}, \textbf{b:} \cite{miranda95}, \textbf{c:} \cite{stanghellini2002}, \textbf{d:} \cite{pottasch2010}, \textbf{e:} \cite{otsuka2020}, \textbf{f:} \cite{sabbadin2005}. 
    \\
    
    }
    \label{tab:boes}
\end{table*}

We carried out deep high-resolution spectroscopy of the 12 PNe from April 2019 to March 2020 using the Bohyunsan Observatory Echelle Spectrograph ({\it BOES}; \citealt{kimkangmin07}) mounted 
on the 1.8 m telescope at the Bohyunsan Optical Astronomy Observatory. {\it BOES} is an optical fiber-fed echelle spectrograph covering a wavelength 
range of 3500--10500 \AA. Raman-scattered \ion{He}{2} features are expected to be broad and weak. To maximize the signal-to-noise ratio, we used the 300 $\mu$m fiber 
which yields a spectral resolution of R ($\equiv\lambda/\Delta\lambda) \sim 30000$, and used 2$\times$2 binning. The total exposure time ranges 
from 600 to 3600 seconds depending on the target brightness. We also observed four spectrophotometric standard stars between our target observations. 

\begin{figure*}
    \centering
    \includegraphics[scale=0.6]{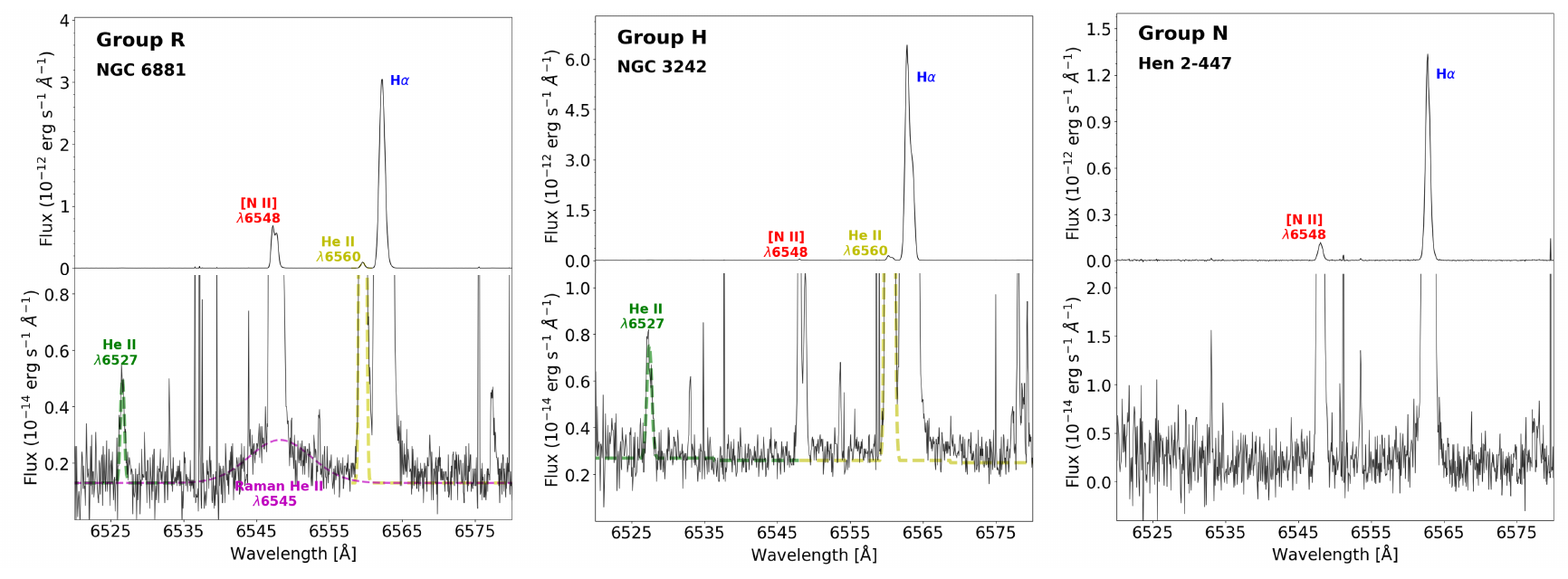}
    \caption{
     Representative spectra near H$\alpha$ of present three distinct groups: Group R, H, and N. Group~R (left, NGC6881) exhibits both Raman-scattered \ion{He}{2}$\lambda6545$ and \ion{He}{2} emission lines. Group~H (center, NGC3242) shows a strong \ion{He}{2} emission line at $6560$ \AA but no Raman-scattered \ion{He}{2}$\lambda6545$. Group~N (right, Hen2-447) shows no detection of Raman-scattered \ion{He}{2}$\lambda6545$ and \ion{He}{2} emission lines in this {\it BOES} observation. The colors of the spectra correspond to the respective lines illustrated in Figure~\ref{fig: schematic_lines}. }  
    \label{fig:represent}
\end{figure*}

The observation log, target list, and their basic information are given in Table~\ref{tab:boes}. 
The data were reduced with the tasks in the {\it ccdred} and {\it echelle} packages in IRAF in a standard way including bias subtraction, extraction of spectral orders, and flat-field corrections. Wavelength and flux calibrations were performed using spectra of the Th--Ar lamp and spectrophotometric standards taken during the observations.
We identify broad wings around [N II]$\lambda6548$ in four targets, which we propose are the Raman-scattered \ion{He}{2}$\lambda6545$. The four objects are NGC~6881, NGC~6886, NGC~6741, and NGC~6884, and the first two of which were previously reported by \citet{choi20b}.

We categorize our 12 target objects into three groups depending on the strength of \ion{He}{2} emission lines and the detection of Raman-scattered \ion{He}{2}$\lambda$6545. 
If no \ion{He}{2} emissions near H$\alpha$ are detected, the objects are classified as Group~N (Non-detection). 
Group R consists of NGC~6881, NGC~6886, NGC~6741, and NGC~6884, where both He II emission at $6527$ \AA \, and 6560 \AA \, and Raman-scattered He II at 6545 \AA \,are detected. The remaining objects that show a strong \ion{He}{2} emission line at 6560 \AA, but no prominent Raman-scattered He II at 6545 \AA \, are classified as Group~H.

%Group R indicates that our failure to detect He II, even in the confirmed He II emitter \citet{tylenda94}, might result from insufficient exposure time during our {\it BOES} observations.

In Figure~\ref{fig:represent}, we show the {\it BOES} spectra of three PNe, each representing one of
the three groups. In the left panel for NGC~6881, broad wings around [\ion{N}{2}]$\lambda$6548
are apparent, which are identified with Raman-scattered \ion{He}{2}$\lambda$6545. It is notable
that the strength of Raman-scattered \ion{He}{2} is comparable to the near \ion{He}{2} emission line at 6527 \AA.
In the middle panel for NGC~3242, no such wings are seen despite the clear detection of \ion{He}{2}$\lambda$6527 and \ion{He}{2}$\lambda$6560. The right panel shows the spectrum of Hen~2-447 with no detectable \ion{He}{2} emission at 6560 \AA.

The evolutionary track for PNe was investigated by many researchers \citep[e.g.,][]{millerbertolami16}. The temperature of the hot central star of a PN may reach $3\times 10^{5}{\rm\ K}$. 
After the central star reaches the highest temperature, 
Raman scattering of \ion{He}{2} will be suppressed due to weakening of
\ion{He}{2}$\lambda$1025. In addition, the neutral material surrounding the central star will be dispersed into the interstellar space, which will lead to further suppression of \ion{He}{2} Raman scattering.

\begin{figure}
\centering
\includegraphics[scale=0.48]{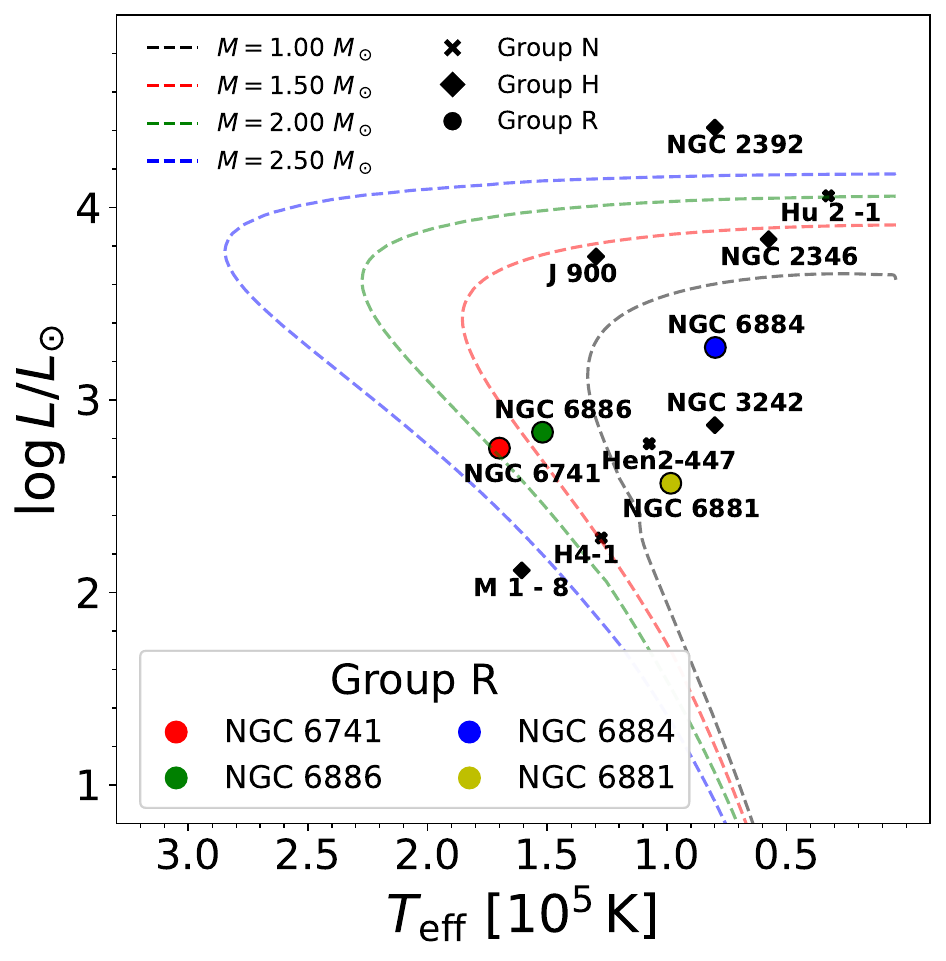}
\caption{Evolutionary tracks between the AGB and the white dwarf phases for stars with the initial masses 
of 1.0, 1.5, 2.0, and 2.5 $M_\odot$ obtained by \cite{millerbertolami16}. 
The points are the temperature and luminosity of the young PNe considered in this paper, as shown in Table~\ref{tab:boes}.
Various marks are used to represent the three groups: Group R (circle), Group H (diamond), and Group N (cross). In particular, the four objects in Group R are shown with colored circles.
}
\label{evolution_track}
\end{figure}

Figure~\ref{evolution_track} shows the stellar evolution tracks between the AGB 
and white dwarf phases for stars with the initial masses of 1.0, 1.5, 2.0, and 2.5 $M_\odot$ obtained by \cite{millerbertolami16}. 
In this figure, the points, with different shapes based on classification, show the positions of the 12 PNe. Here, the temperature and luminosity are taken 
from the literature shown in Table ~\ref{tab:boes}. In particular, the filled circles correspond to the four PNe showing Raman-scattered \ion{He}{2} feature at 6545~\AA.   

%Specific consideration 3 
The 12 PNe observed with {\it BOES} are situated in a region characterized by very high temperatures, exceeding \( 3 \times 10^4 \, \mathrm{K} \), and luminosities greater than \( 10^3 L_\odot \). Notably, four of these objects exhibiting Raman-scattered \ion{He}{2}$\lambda$6545 are closely interspersed with the other eight near the turning point. This raises the intriguing question of whether these four PNe in Group~R can be distinctly identified using other independent observational factors.
A preliminary inspection suggests a systematic trend of larger Balmer decrement in Group R objects. However, further studies targeting a larger sample of \ion{He}{2} emitting PNe are necessary to draw definitive conclusions.

\section{Line Profile Analysis}

\subsection{Gaussian fitting}
\begin{table*}
\caption{Line fit parameters of the four objects in Group~R} 
\setlength{\tabcolsep}{6pt}
\renewcommand{\arraystretch}{1.1}
\begin{threeparttable}
\centering
\begin{tabular}{c|c|c|c|c|c}
    \hline
  \multirow{2}{*}{\textbf{Line}}& \multirow{2}{*}{\textbf{Gaussian Fit Parameter}}& \multirow{2}{*}{\textbf{NGC6881}}& \multirow{2}{*}{\textbf{NGC6886}}& \multirow{2}{*}{\textbf{NGC6741}}&\multirow{2}{*}{\textbf{NGC6884}}\\
           &&  &  &  & \\
    \hline
 He II 6527& \multirow{2}{*}{$\lambda_{c} $}
& 6526.58& 6526.83& 6527.46&6526.11\\
 He II 6560& \multirow{2}{*}{(\AA)}& 6559.58& 6559.86& 6560.46&6559.10\\
 Raman He II 6545& & 6548.23& 6549.00& 6548.70&6548.06\\
 \hline
 $\rm H\alpha$&& $2.9 \, \times \, 10^{-12} $& $1.7 \, \times \, 10^{-11}$& $2.5 \, \times \, 10^{-11}$&$4.0 \, \times \, 10^{-11}$\\
            He II 6527& $F_{\rm tot} $&   $2.4 \, \times \,10^{-15}$&   $1.5\, \times \,10^{-14}$&   $2.9 \, \times \, 10^{-14}$&  $2.1\, \times \,10^{-14}$\\
            He II 6560&$(\rm erg\ s^{-1}\ cm^{-2})$&   $6.0 \, \times \, 10^{-14}$&   $3.5 \, \times \,10^{-13}$&   $4.3\, \times \,10^{-13}$&  $4.8\, \times \, 10^{-13}$\\
           Raman He II 6545&&   $1.6 \, \times \,10^{-14}$&   $6.3 \, \times 10^{-14}$&   $6.6\, \times \,10^{-14}$& $2.7\, \times \, 10^{-14}$\\

    \hline         
 He II 6527&\multirow{2}{*}{${v_{\rm G}} $}&  11&  14&  15& 17\\
  He II 6560& 
\multirow{2}{*}{($\rm km \  s^{-1})$}& 11& 14& 15&17\\
 {Raman He II 6545}& & 196& 173& 194&123\\
  \hline
  
  \multicolumn{2}{c|}{${v_{\rm G}^{\rm Ram}}$\tnote{*}}& 31& 27& 30&19\\
   \hline
  \multicolumn{2}{c|}{$F_{6560} / F_{6527}$}& 25& 23& 15&23\\
  \hline
 \multicolumn{2}{c|}{$F_{6560} / F_{\rm H\alpha}$}& 0.020& 0.021& 0.017&0.012\\
   \hline
  \multicolumn{2}{c|}{$\Delta V_c$}& \multirow{2}{*}{+29}& \multirow{2}{*}{+33}& \multirow{2}{*}{+26}&\multirow{2}{*}{+31}\\
  \multicolumn{2}{c|}{$(\rm km \  s^{-1})$}& & & &\\
   \hline
\end{tabular}
\begin{tablenotes}
\item [*]
The velocity width $v^{\rm Ram}_G$ of Raman-scattered \ion{He}{2}$\lambda$6545 in the parent velocity space of \ion{He}{2}$\lambda$1025 using Equation~(\ref{eq:ram_width_6.4}).  
\end{tablenotes}
\end{threeparttable}
\label{tab:parameters}

\end{table*}

\begin{figure*}
    \centering
     \includegraphics[scale=0.35]{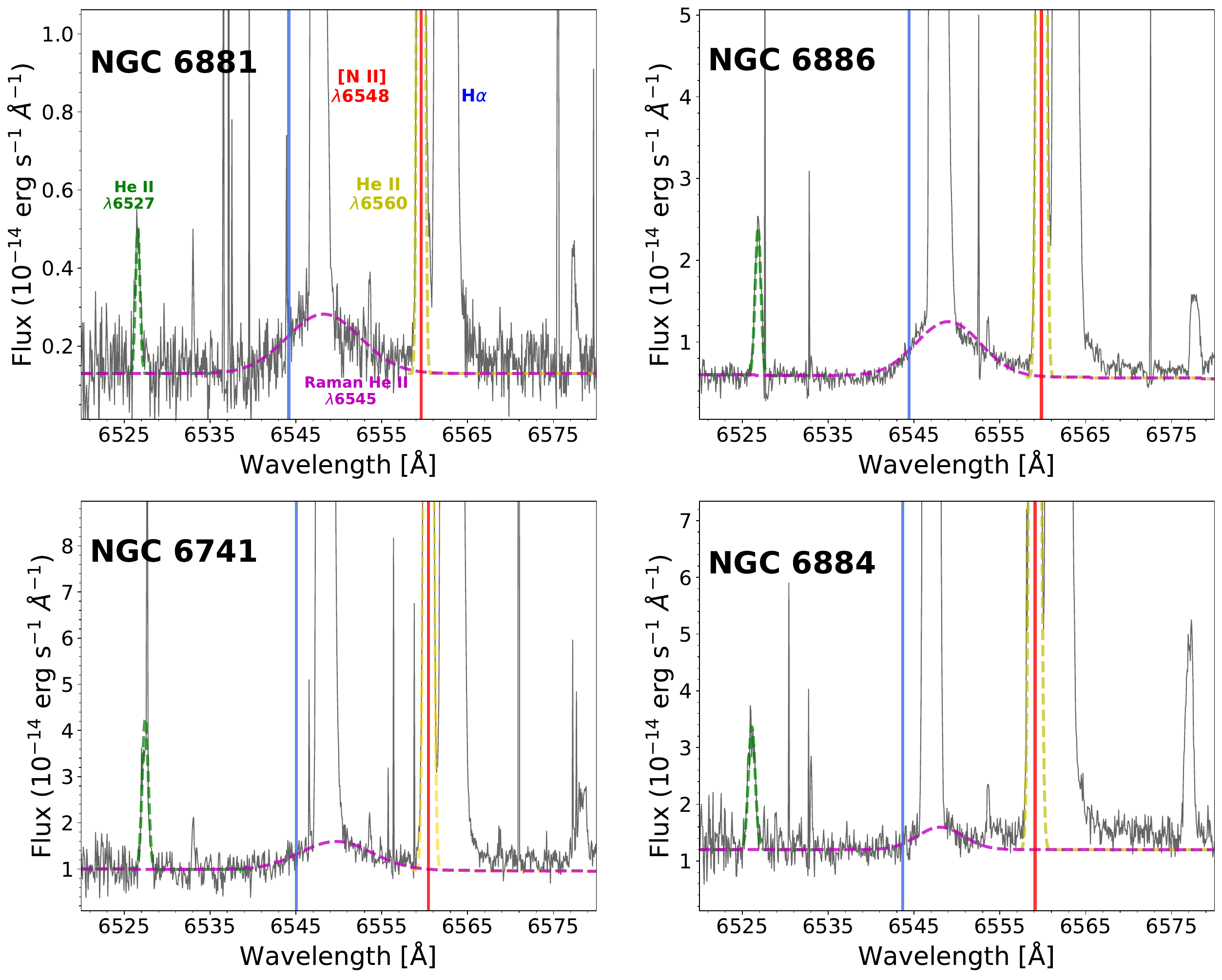}    
    \caption{{\it BOES} spectra of NGC~6881, NGC~6886, NGC~6741, and NGC~6884 in Group R. These PNe have a relatively weak but broad Raman \ion{He}{2} $\lambda$6545 line blended with the strong [\ion{N}{2}]$\lambda$6548 line, along with clear \ion{He}{2} $\lambda$6527 and $\lambda$6560 emission lines. The solid line represents the observed data, while the colored dashed lines denote the Gaussian fit. Colors green, yellow, and purple correspond to \ion{He}{2}$\lambda6527$, \ion{He}{2}$\lambda6560$, and Raman-scattered \ion{He}{2}$\lambda6545$, respectively. Two vertical colored lines are drawn, where the red lines indicate the observed center of \ion{He}{2}$\lambda$6560, and the blue lines mark the expected atomic line center of Raman-scattered \ion{He}{2}$\lambda$6545 based on the center line of \ion{He}{2}$\lambda6560$.}  
    \label{fig:Group_R}
\end{figure*}

To detect and analyze Raman-scattered \ion{He}{2}, we perform a line fitting analysis for every visible emission line in the wavelength window of 6500--6600\AA\ (i.e., \ion{He}{2}$\lambda\lambda$6527, 6560 and H$\alpha$). 
The analysis is carried out in three steps. First, we fit and subtract local continuum across 6500--6600\AA, determined from the emission-free windows of 6530--6540\AA\ and 6625--6635\AA.
Secondly, we fit the emission lines in the continuum-subtracted spectrum. For \ion{He}{2}$\lambda$6527 and \ion{He}{2}$\lambda$6560, H$\alpha$, which exhibit single-peak structures, we use a single Gaussian function given as
\begin{equation}
    F(\lambda)=F_0 \exp{[-(\lambda-\lambda_c)^2/2\Delta\lambda^2]},
\label{eq:gauss_fit}
\end{equation}
where $F_0$, $\lambda_c$ and $\Delta\lambda$
 are the peak flux density, the observed line center and the width of the Gaussian function, respectively. In our line fitting analysis, we tie the kinematics between \ion{He}{2}$\lambda$6527 and \ion{He}{2}$\lambda$6560.
The total line flux $F_{\rm tot}$ is given by
 
\begin{equation}
    F_{\rm tot} = \sqrt{2\pi} F_0 \Delta\lambda.
    \label{eq:tot_flx}
\end{equation}
In contrast, for [\ion{N}{2}]$\lambda\lambda$6548, 6583, we use a double Gaussian function due to their prominent double-peak features. In the line fit analysis, the kinematics of these two lines is tied with the flux ratio fixed to be 1:3 \citep{osterbrock89}.
The best-fit parameters are determined by minimizing the sum of the chi-squared values. We correct the line widths by accounting for the instrumental resolution of {\it BOES} ($R \sim 30,000$).

Lastly, we mask [\ion{N}{2}]$\lambda$6548 and fit Raman-scattered \ion{He}{2}$\lambda$6545 separately. This is because of the severe blending between Raman-scattered \ion{He}{2}$\lambda$6545 and [\ion{N}{2}]$\lambda$6548.
Note that we attempted to fit the two emission lines simultaneously but found that 
the fitting of Raman-scattered \ion{He}{2}$\lambda$6545 is heavily affected by the fitting result of  [\ion{N}{2}]$\lambda$6548 due to its much weaker line strength compared to [\ion{N}{2}]$\lambda$6548 (i.e., $\sim$1/500).
Therefore, we decide to fit Raman-scattered \ion{He}{2}$\lambda$6545 separately to avoid the contamination
from [\ion{N}{2}]$\lambda$6548.
In view of the fact that [\ion{N}{2}]$\lambda$6548 exhibits a double-peak structure 
in the four objects having Raman-scattered \ion{He}{2}, 
the masking method is applied for both blue and red, within a $5\sigma$ threshold. 
After fitting, we determine the best-fit parameters of \ion{He}{2}$\lambda$6545 in the same manner as for the other lines.

In Figure~\ref{fig:Group_R}, the observed data
are shown by black solid lines. The green and yellow dashed lines show the fitting Gaussians for \ion{He}{2}$\lambda$6527 and \ion{He}{2}$\lambda$6560, respectively, while the purple dashed line is the Raman-scattered \ion{He}{2}$\lambda$6545. 
It is noted that Raman-scattered \ion{He}{2} features are significantly broader than other \ion{He}{2} emission lines.

In Table~\ref{tab:parameters}, we provide our fitting results for the four PNe in Group R.
Here, instead of $\Delta\lambda$ we present the velocity width of the observed emission lines defined by
\begin{equation}
    v_G =  \left( {\Delta\lambda \over \lambda_c} \right) c.
\label{eq:ram_width}
\end{equation}
In the case of Raman-scattered \ion{He}{2}$\lambda$6545, the velocity width is computed in the parent velocity space so that
\begin{equation}
    v_G^{\rm Ram} = {1 \over 6.4} \left( {\Delta\lambda \over \lambda_c} \right) c,
\label{eq:ram_width_6.4}
\end{equation}
where the factor 6.4 is the ratio of the frequencies of the incident and 
Raman-scattered photons.

\subsection{Line Center Shift of Raman-scattered Line}

In Figure~\ref{fig:Group_R}, two vertical lines are drawn, where the red lines indicate the observed
line center of \ion{He}{2}$\lambda6560$. 
From the atomic physics of  \ion{He}{2} Raman scattering, Raman-scattered \ion{He}{2}$\lambda6545$ is supposed to be formed at $\Delta v= -704 ~ \rm km\,s^{-1}$ from the line center of \ion{He}{2}$\lambda6560$.  The atomic line centers, $\lambda^{\rm atomic}_{6545}$ and $\lambda^{\rm atomic}_{6560}$ adopted in this work are  6544.70 \AA\  and  6560.10  \AA, respectively (\citeauthor{hyung04} \citeyear{hyung04}; \citeauthor{lee06} \citeyear{lee06}; \citeauthor{chang23} \citeyear{chang23}). 
The blue vertical lines show the expected atomic center of Raman-scattered \ion{He}{2}$\lambda6545$  based on the position marked by the blue lines.

It is clearly noted that the observed Raman-scattered \ion{He}{2}$\lambda6545$ features are significantly redshifted with respect to the \ion{He}{2} emission region, which we attribute to the expansion of the \ion{H}{1} component with respect to the \ion{He}{2}
emission region. In order to determine the expansion speed, we may note that \ion{He}{2} optical emission lines serve as velocity references for this purpose.
The relative velocity $\Delta V_c$ between the \ion{H}{1} and \ion{He}{2} regions is determined by
\begin{equation}
   {\Delta V_c \over c} = \left(\frac{\lambda^{\rm atom}_{1025}}{\lambda^{\rm atom}_{6545}}\right) \ \frac{(\lambda^{\rm atom}_{6560} - \lambda^{\rm atom}_{6545}) - (\lambda^{\rm obs}_{6560} - \lambda^{\rm obs}_{6545}) }{\lambda^{\rm atom}_{6545}} ,
    \label{equ:relative}
\end{equation}
 where $\lambda^{\rm obs}$ and $\lambda^{\rm atom}$  are the wavelengths of the observed line center and
 the atomic line center, respectively.
 The velocity offset $\Delta V_c$ yields the same result when $\lambda_{6560}$ is replaced with $\lambda_{6527}$.

It is interesting that the relative velocities between \ion{H}{1} and \ion{He}{2} regions
of the four objects are found in the considerably narrow range of $26-33{\rm\ km\ s^{-1}}$, as shown in Table~\ref{tab:parameters}.
This velocity range is consistent with the velocity scale expected of the slow stellar
wind from stars on the AGB track \citep[e.g.,][]{hofner18}. 
\cite{taylor90}  carried out 21 cm observations to search the \ion{H}{1} component 
in young PNe. In the case of NGC~6886, they found the expansion velocity of the \ion{H}{1}
component is $v_{\rm exp}=22.8 \pm 1.2 {\rm\ km\ s^{-1}} $.

\section{Photoionization \& Raman Conversion Efficiency}

To trace the \ion{H}{1} mass in the vicinity of the \ion{He}{2} emission region, it is essential to estimate the Raman conversion efficiency (RCE) and perform radiative transfer modeling.
The Raman conversion efficiency is defined as
\begin{equation}\label{eq:RCE1} {RCE}_{6545} = {{\Phi{6545}} \over {\Phi_{1025}}} , \end{equation}
where $\Phi_{1025}$ and $\Phi_{6545}$ are the total number fluxes of \ion{He}{2} $\lambda$1025 and Raman-scattered \ion{He}{2} $\lambda$6545, respectively.
However, the interstellar extinction is extremely heavy near the Lyman series of \ion{H}{1}, making direct measurement of $\Phi_{1025}$ infeasible. One way to address this challenge is to use photoionization modeling, which allows us to deduce the flux ratio of \ion{He}{2}$\lambda$1025 to \ion{He}{2}$\lambda$6560.
Using the flux ratio, Equation~(\ref{eq:RCE1}) can be rewritten as
\begin{equation}\label{eq:RCE2} {RCE}_{6545} = \left( \lambda{0, 6545} \over \lambda_{0, 1025} \right) \left( F_{6560} \over F_{1025} \right) {\left({F_{6545}} \over {F_{6560} } \right)} , \end{equation}
where $\lambda_{0, 1025}$ and $\lambda_{0, 6545}$ are the central wavelengths of \ion{He}{2} $\lambda$1025 and Raman-scattered \ion{He}{2} $\lambda$6545. This formulation relates the Raman conversion efficiency to observable fluxes and wavelengths, enabling indirect estimation even under conditions of strong interstellar extinction.

% Using the deduced $F_{1025}$,
We use the Monte Carlo radiative transfer code \texttt{STaRS} \cite{chang20} with a simple scattering geometry to constrain the \ion{H}{1} distribution that is consistent with the Raman conversion efficiency deduced from photoionization modeling.
In Sections~\ref{sec:photoinoization} and \ref{sec:Geometry}, we present the photoinization computations and radiative transfer modeling, respectively.

% In Section~\ref{sec:RCE}, we show simulated results $RCE_{6545}$ and compare them with observations.

%The photoionization model is presented in Section~\ref{sec:photoinoization}, followed by descriptions of the scattering geometry and the Raman conversion efficiency in Sections~\ref{sec:Geometry} and \ref{sec:RCE}, respectively.

\begin{figure*}
    \centering
    \includegraphics[scale=0.45]{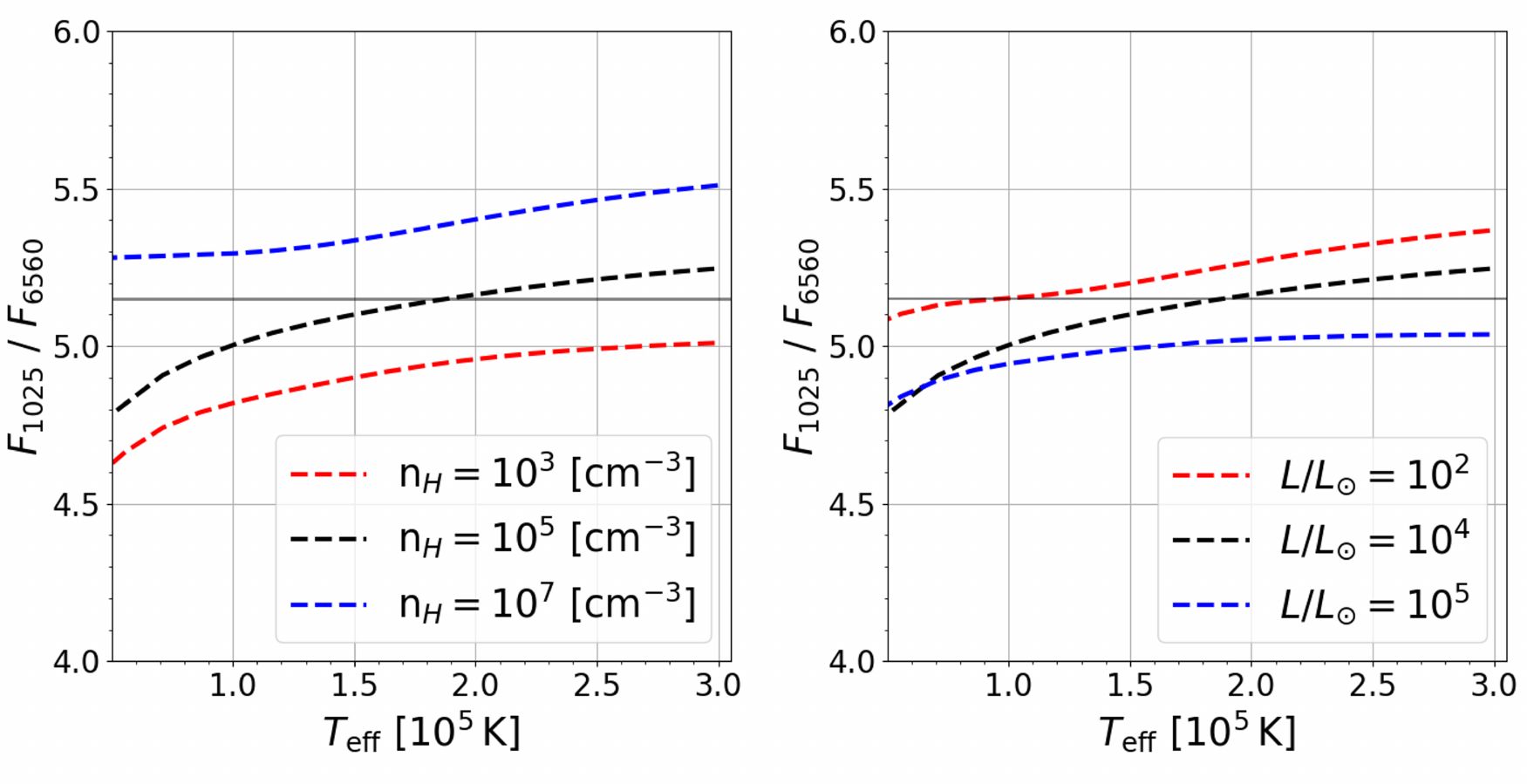} 
    \caption{The flux ratios $F_{1025}/F_{6560}$ for various values of $T_{\rm eff}$ obtained from our photoionization modeling using \texttt{CLOUDY}.
    The left panel displays the flux ratios for three values of $n_{\rm H}$ = $10^3$ (red), $10^5$ (black), and $10^7\, \rm cm^{-3}$ (blue) with fixed $L_* = 10^4\ L_{\odot}$.
    In the right panel, the flux ratios are shown for three different values of the luminosity of the central star, $L_* = 10^2\  L_\odot$ (red), $L_* = 10^2 \ L_\odot$ (black), and $L_* = 10^5\ L_\odot$ (blue) with the number density  fixed at $n_{\rm H} = 10^5 \, \rm cm^{-3}$. } 
    \label{fig:cloudy}
\end{figure*}

\subsection{Photoionization Model of He~II Emission}\label{sec:photoinoization}

The publicly available photoionization modeling code \texttt{CLOUDY} \citep{ferland17} is widely used to investigate the physical parameters of the observed spectra of PNe illuminated by their hot central stars \citep{barria2018, otsuka2020, rebeca2022, gomez2024}.
Young PNe with detected Raman-scattered \ion{He}{2} lines exhibit complex nebular morphologies, ranging from bipolar to prolate ellipsoidal shapes. This suggests that the \ion{He}{2} emission regions may have intricate geometrical and kinematical structures \citep[e.g.,][]{santander_garcia17, balick23}.

In this work, we utilize \texttt{CLOUDY} with a simplified geometry to investigate the dependence of the \ion{He}{2} flux ratio $F_{6560}/F_{1025}$ on the physical conditions of the source and surrounding gas. A spherical nebula with uniform density, photoionized by black body radiation, is assumed. To further simplify the model, solar abundance is adopted.
Three main parameters are considered: the hydrogen number density of the emission nebula, $n_H = 10^{3-7} \, \rm cm^{-3}$; the effective temperature of the central star, $T_{\rm eff} = 50,000 - 300,000 \, \rm K$; and the stellar luminosity, $L_*/L_{\odot} = 10^{2-5}$.

In Figure~\ref{fig:cloudy}, we present the computed \ion{He}{2} flux ratio $F_{1025}/F_{6560}$ as a function of $T_{\rm eff}$. The left panel illustrates how the flux ratios vary with the hydrogen number density, $n_{\rm H}$. In contrast, the right panel displays the flux ratios for different stellar luminosities, $L_*$. 
Across the range considered in our study, $F_{1025}/F_{6560}$ increases from 4.6 to 5.5.

Consequently, we adopt the flux ratio $F_{1025}/F_{6560} = 5.15$, which represents the average value for $T_{\rm eff}$ at $n_{\rm H} = 10^5 \, {\rm cm^{-3}}$ and $L_{*} = 10^4 L_{\odot}$, to compute the Raman Conversion Efficiency $RCE_{6545}$.
With this flux ratio, Equation~(\ref{eq:RCE2}) simplifies to
\begin{equation} RCE_{6545} = 1.24 \left(\frac{F_{6545}}{F_{6560}} \right). \label{step2}
\end{equation}
Using the observed values listed in Table~\ref{tab:parameters}, we calculate the Raman conversion efficiencies for \ion{He}{2}$\lambda$1025 as 0.39, 0.24, 0.21, and 0.07 for NGC~6881, NGC~6886, NGC~6741, and NGC~6884, respectively.

\subsection{Radiative Transfer Modeling}\label{sec:Geometry}

To investigate the distribution of \ion{H}{1} consistent with the observed $RCE_{6545}$, we perform Monte Carlo simulations of the formation of Raman-scattered \ion{He}{2}. For this purpose, we use the \texttt{STaRS} code developed by \cite{chang20}, adopting the same scattering geometry as described in \cite{choi20a}.

\begin{figure}
    \centering
    \includegraphics[width=0.50\textwidth]{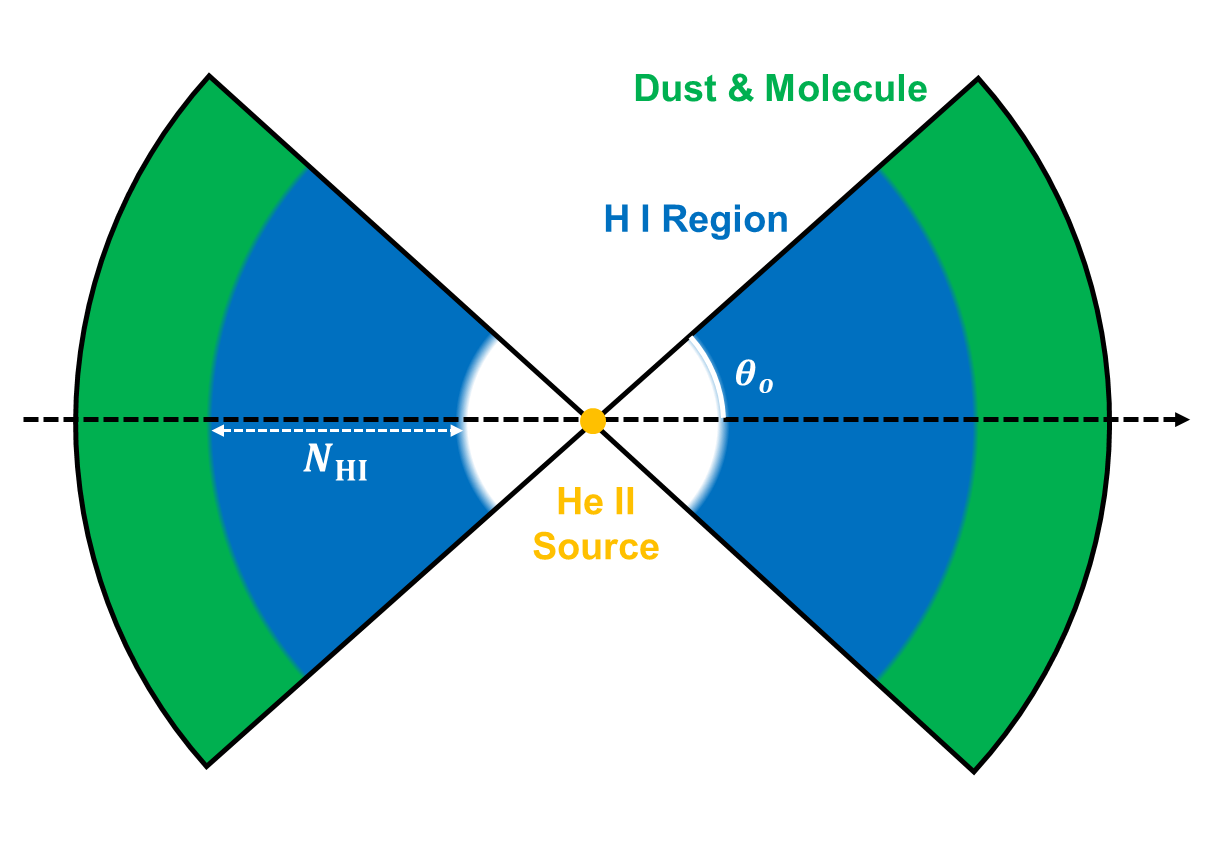}
    \caption{
     A schematic illustration of the model geometry composed of a point-like He II emission source (orange), an \ion{H}{1} region (blue) characterized by \ion{H}{1} column density $N_{\rm HI}$ and the half opening angle $\theta_{\rm o}$, and dusty molecular region (green). 
    }
    \label{fig:geometry}
\end{figure}

Figure~\ref{fig:geometry} describes a schematic illustration of the model geometry. In this scattering geometry, the \ion{H}{1} region (blue) forms an open shell surrounding the central star, which is represented as a point-like \ion{He}{2} emission region. The \ion{H}{1} disk is defined by the half-opening angle $\theta_{\rm o}$ and the \ion{H}{1} column density $N_{\rm HI}$ \citep{chang23}.
The dusty molecular region (green) is located outside the \ion{H}{1} region. This structure is motivated by the observation of Raman-scattered \ion{He}{2} lines in NGC~6302 \citep{chang23}, which is significantly obscured by dust \citep{kastner22}.

As long as the dusty region lies outside the \ion{H}{1} region, \ion{He}{2}$\lambda1025$ photons are not affected by dust extinction before being Raman-scattered within the \ion{H}{1} region. Additionally, both Raman-scattered \ion{He}{2}$\lambda6545$ photons and optical \ion{He}{2}$\lambda6560$ photons undergo nearly identical dust extinction due to their minimal wavelength difference. Consequently, the flux ratio of the observed \ion{He}{2}$\lambda6560$ to Raman-scattered \ion{He}{2}$\lambda6545$, as defined in Equation~(\ref{eq:RCE2}), determines the Raman conversion efficiency. This ratio remains unaffected by dust extinction because both lines are influenced equally.

In this simulation's scattering geometry, the \ion{H}{1} medium expands away from the \ion{He}{2} emission region with a velocity \( v_{\rm exp} \). \cite{choi20a} demonstrated that the Raman conversion efficiency depends on the expansion velocity (see Figure~6). Additionally, \cite{chang23} presented model spectra of NGC~6302 based on Monte Carlo simulations for various expansion velocities of the Raman-scattered \ion{He}{2} lines. Their findings suggest that a static \ion{H}{1} region is inconsistent with the observed spectrum, strongly indicating the need to incorporate the expanding motion.

\begin{figure*}
    \centering
    \includegraphics[width=1.0\textwidth]{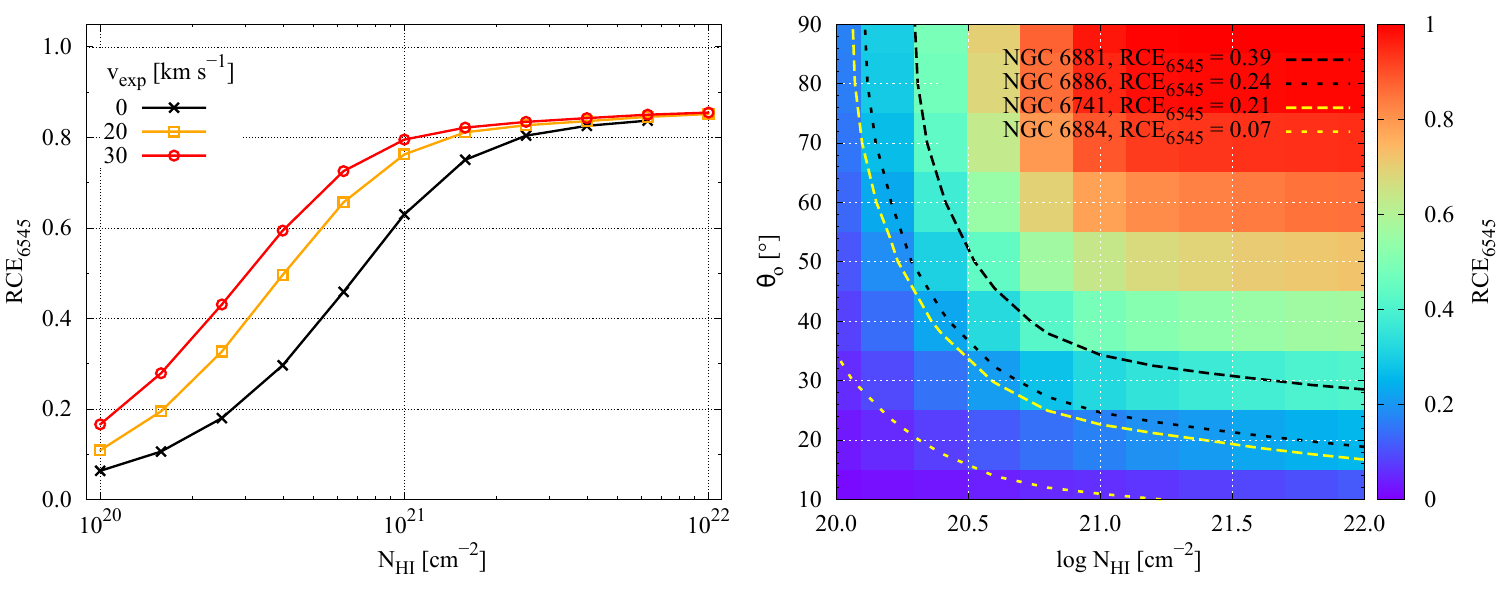}
    \caption{ The Raman conversion efficiency, \( RCE_{6545} \), is computed from Monte Carlo simulations using the \texttt{STaRS} code, assuming an H~I disk geometry surrounding the central \ion{He}{2} source.  The left panel shows \( RCE \) values for \( v_{\rm exp} = 0 \) (black), 20 (orange), and 30 \( \mathrm{km\,s^{-1}} \) (red) as a function of \( N_{\rm HI} \) with \( \theta_{\rm o} = 60^\circ \). 
    The right panel displays a map of \( RCE \) in the \( N_{\rm HI}-\theta_{\rm o} \) plane for \( v_{\rm exp} = 30~\mathrm{km\,s^{-1}} \). The contours represent the observed \( RCE_{6545} \) values for NGC~6881 (black dashed), NGC~6886 (black dotted), NGC~6741 (yellow dashed), and NGC~6884 (yellow dotted).}
    \label{fig:Group_R_6545}
\end{figure*}

Our Monte Carlo simulation results are shown in Figure~\ref{fig:Group_R_6545}.  
The left panel presents the \( RCE_{6545} \) values for $\theta_{\rm o} = 60^\circ$ at three different expansion velocities: \( v_{\rm exp} = 0 \) (black), 20 (orange), and 30 \( \mathrm{km\,s^{-1}} \) (red). The right panel displays the \( RCE_{6545} \) values computed over a range of \( \theta_{\rm o} \) and \( N_{\rm HI} \) for \( v_{\rm exp} = 30~\mathrm{km\,s^{-1}} \), corresponding to the \( \Delta V_c \) values for four objects in Group R (cf. Table~\ref{tab:parameters}). The contours (colored dashed lines) represent the \( RCE_{6545} \) values estimated for the four objects.  
For \( \theta_{\rm o} = 90^\circ \) and \( N_{\rm HI} > 10^{21} \, \mathrm{cm^{-2}} \), all \ion{He}{2} \( \lambda 1025 \) photons undergo Raman scattering, resulting in \( RCE_{6545} = 1 \).

From the results shown in Figure~\ref{fig:Group_R_6545}, and assuming a scattering geometry with a column density of \( N_{\rm HI} = 10^{20.5}\ {\rm cm^{-2}} \), we estimate the \ion{H}{1} masses to be \( 6 \times 10^{-2} M_{\odot} \), \( 5 \times 10^{-2} M_{\odot} \), \( 5 \times 10^{-2} M_{\odot} \), and \( 2 \times 10^{-2} M_{\odot} \) for NGC~6881, NGC~6886, NGC~6741, and NGC~6884, respectively.  As \( N_{\rm HI} \) and \( \theta_{\rm o} \) increase, multiple scattering effects become significant, leading to a more complex relationship between \( RCE_{6545} \) and the \ion{H}{1} mass.

\cite{choi20b} set the column densities of NGC~6886 and NGC~6881 as \( N_{\rm HI} = 5 \times 10^{20} \, \mathrm{cm^{-2}} \) and \( N_{\rm HI} = 3 \times 10^{20} \, \mathrm{cm^{-2}} \), respectively. The \ion{H}{1} masses of NGC~6881 and NGC~6886 are estimated to be \( 4 \times 10^{-2} \, M_{\odot} \) and \( 3 \times 10^{-2} \, M_{\odot} \), respectively. 
The differences in mass estimates are primarily attributed to the adopted values of \( N_{\rm HI} \). Additionally, they applied case B recombination theory to deduce \ion{He}{2}$\lambda$1025 and performed line fitting using simulated profiles instead of single Gaussian functions. Improving the \ion{H}{1} mass estimates may require more intensive investigations with sophisticated scattering geometry.

The values of \( RCE_{6545} \) for NGC~6741 and NGC~6884 are lower than those for NGC~6881 and NGC~6886, where the detection of Raman-scattered \ion{He}{2} was reported by \cite{choi20b}. Particularly for NGC~6884, \( RCE_{6545} \) is less than 10 percent, making the Raman-scattered \ion{He}{2} emission very weak. Additionally, the line width is considerably narrower compared to the other three objects.  
\cite{choi20a} demonstrated that Raman-scattered \ion{He}{2} lines can become significantly broadened due to multiple scattering in a very thick \ion{H}{1} medium. The low \( RCE_{6545} \) and narrow line width observed for NGC~6884 are consistent with a smaller \ion{H}{1} mass compared to the other three objects.

\section{Discussion}

\subsection{Evolution of PNe and Raman Spectroscopy}

\begin{figure*}
    \centering
    \includegraphics[scale=0.4]{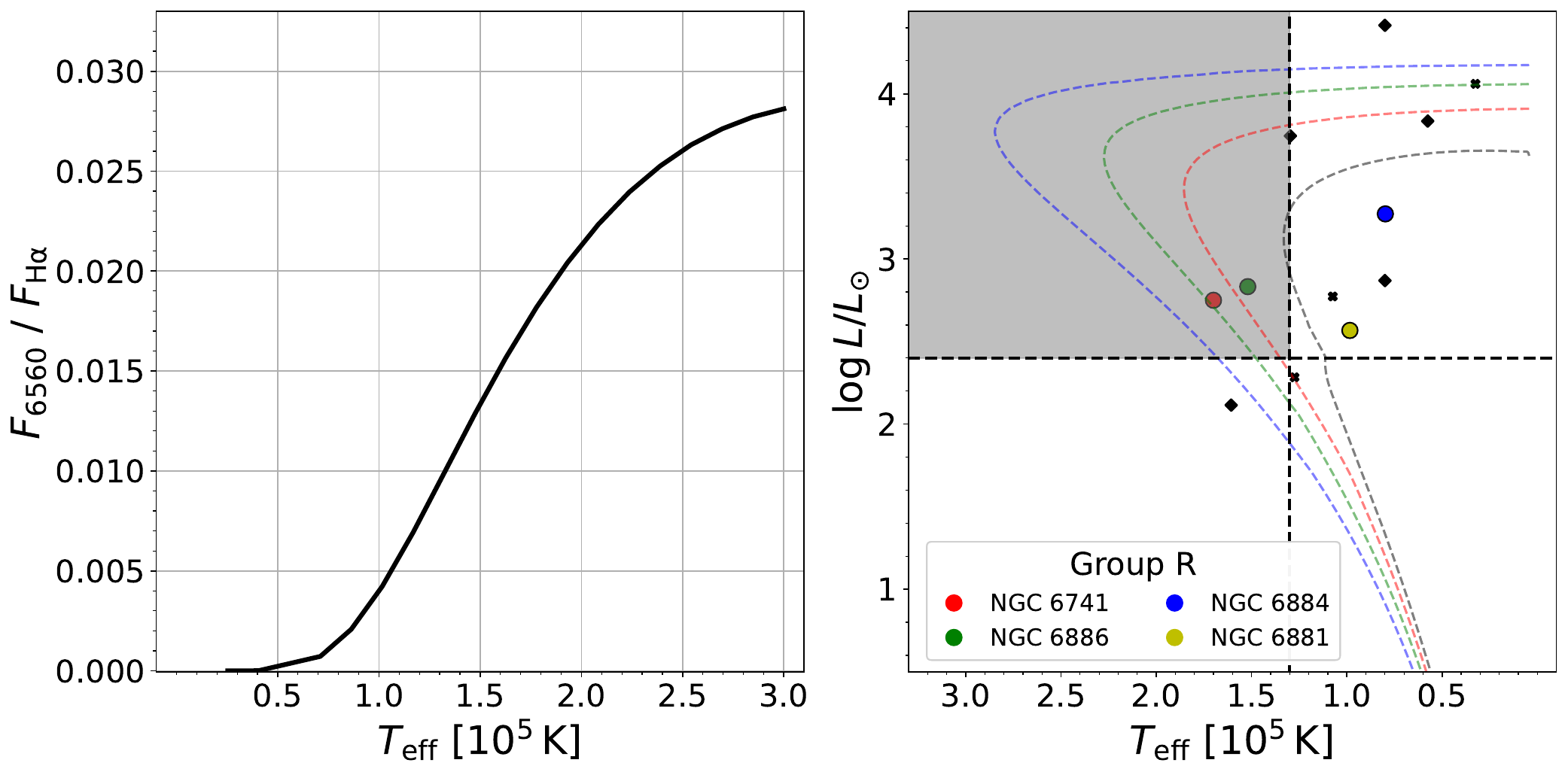}
    \caption{The flux ratio $F_{6560} / F_{\rm H\alpha}$ computed using \texttt{CLOUDY} for an emission nebula characterized by our representative values $n_{H} = 10^5\,\rm cm^{-3}$ and $L = 10^4 L_{\odot}$ for a range of $T_{\rm eff}$ (left), and
    the evolutionary tracks of planetary nebulae shown in Figure~\ref{evolution_track} (right).  
In the right panel, the vertical dotted line represents the temperature $T_{\rm eff}=1.3\times10^5{\rm\ K}$ of
the photoionization source that is required for $F_{6560}/F_{\rm H\alpha}>0.01$.
The horizontal dotted line represents the lower bound for the luminosity of the PNe
with Raman-scattered \ion{He}{2} in our {\it BOES} data.}
    \label{fig:evol_pn}
\end{figure*}

The Raman scattering process of \ion{He}{2} in PNe requires a thick \ion{H}{1} region illuminated by a strong \ion{He}{2} source. According to our photoionization model calculations using \texttt{CLOUDY} for a spherical nebula ionized by a central hot star, the strength of \ion{He}{2} emission relative to \ion{H}{1} is primarily determined by the temperature of the central ionizing source, with little dependence on the source's luminosity or the nebula's density.  
In the left panel of Figure~\ref{fig:evol_pn}, we present the flux ratio \( F_{6560} / F_{\rm H\alpha} \) for an emission nebula characterized by representative values of \( n_{\rm H} = 10^5 \,\mathrm{cm^{-3}} \) and \( L = 10^4 L_{\odot} \), while varying \( T_{\rm eff} \) in the range \( 3.0 \times 10^4 \,\mathrm{K} < T_{\rm eff} < 3.0 \times 10^5 \,\mathrm{K} \). The flux ratio is less than 0.001 for \( T_{\rm eff} = 50,000 \,\mathrm{K} \), while for \( T_{\rm eff} = 200,000 \,\mathrm{K} \), it increases to 0.02, which is typical of the \ion{He}{2} emitting PNe in our sample (cf. Table~\ref{tab:parameters}).

The right panel of Figure~\ref{fig:evol_pn} displays the evolutionary tracks of PNe, as illustrated in Figure~\ref{evolution_track}. The shaded region, bounded by a vertical dotted line and a horizontal dotted line, approximately represents the evolutionary stage where Raman He II spectroscopy may play a significant role. The horizontal dotted line marks the lower luminosity bound of the PNe with Raman-scattered \ion{He}{2} detected in our {\it BOES} data.  
As a PN evolves and crosses this horizontal line, Raman scattering of \ion{He}{2} is severely suppressed due to the dilution of neutral material into interstellar space, combined with the decreasing far-UV \ion{He}{2} emission as the central star cools. The vertical dotted line indicates the temperature \( T_{\rm eff} \) of the photoionizing source that results in a flux ratio \( F_{6560}/F_{\rm H\alpha} > 0.01 \). This implies that strong \ion{He}{2}-emitting PNe likely host central stars with \( T_{\rm eff} > 1.3 \times 10^5 \, \mathrm{K} \).  
It is noted that two PNe—NGC~6881 (yellow) and NGC~6884 (blue)—are located outside the shaded region. However, given the strength of \ion{He}{2} emission in these objects, it is highly likely that their temperatures are significantly underestimated. A detailed discussion of the uncertainties in measuring the temperature and luminosity of these objects is beyond the scope of this article.

Considering the significance of their evolutionary status near the turning point in the H-R diagram, a refined estimate of the population will offer crucial insights into the final stages of stellar evolution. A systematic spectroscopic survey of Raman-scattered \ion{He}{2} lines could provide valuable clues about the photoionization and photodissociation processes occurring during the early stages of PN evolution.

\subsection{Raman Wings around Balmer Lines}

Another intriguing feature associated with Raman scattering by atomic hydrogen is the broad wings often observed in the Balmer emission lines of PNe. Notable examples are found in young PNe, including M2-9, IC~4997, IC~5117, and M3-27 \citep[e.g.,][]{balick89, lee00, arrieta03, ruizescobedo24}. These wings can naturally form in a fast, hot stellar wind emanating from the central star.  
For instance, the far-UV spectrum of NGC~6543 exhibits prominent P~Cygni profiles in the resonance \ion{O}{6}$\lambda\lambda$1032 and 1038 lines, indicating the presence of a fast stellar wind with a velocity exceeding \( 10^3 \, \mathrm{km \, s^{-1}} \) \citep{perinotto89}.

Raman scattering by atomic hydrogen of far-UV radiation near the Lyman lines produces optical photons around the Balmer lines, leading to the formation of wings \citep{nussbaumer89, chang18}. Broad Balmer wings are commonly observed in symbiotic stars, which also exhibit Raman-scattered \ion{O}{6} and \ion{He}{2} features. This suggests that the far-UV continuum near the Lyman series plays a significant role in the formation of Balmer wings in these objects.

\section{Summary and Future Works}

We conducted deep, high-resolution spectroscopy of 12 young PNe using {\it BOES} and identified the Raman-scattered \ion{He}{2} feature near 6545 \AA\ in two additional young planetary nebulae, NGC~6741 and NGC~6884. These discoveries add to the previously reported detections in NGC~6881 and NGC~6886 by \cite{choi20b}.

The strength of Raman-scattered \ion{He}{2} depends on both the incident \ion{He}{2} emission and the amount of \ion{H}{1} surrounding the \ion{He}{2} region. As the PN evolves, the amount of \ion{H}{1} is expected to decrease as the entire nebula expands and the central star cools. Similarly, the relative strength of \ion{He}{2} emission compared to \ion{H}{1} is also expected to vary throughout the evolution of the PN.

The spectroscopic analysis of Raman-scattered \ion{He}{2}$\lambda$6545 is challenging due to severe blending with the strong [\ion{N}{2}]$\lambda$6548 line. In the cases of NGC~7027, NGC~6302, and IC~5117, Raman-scattered \ion{He}{2}$\lambda$4850 has been successfully detected \citep{pequignot97, groves02, chang23}.  
Unlike \ion{He}{2}$\lambda$6545, Raman-scattered \ion{He}{2}$\lambda$4850 is free from blending with other spectral features, allowing for more sophisticated theoretical modeling. However, our {\it BOES} observations are not deep enough to detect Raman-scattered \ion{He}{2}$\lambda$4850.

It is noteworthy that the detection of Raman-scattered \ion{He}{2} has predominantly been limited to objects in the northern sky, with NGC~6302 being the only exception. This underscores the importance of conducting spectroscopic surveys aimed at identifying Raman-scattered \ion{He}{2} features in the southern sky. By adopting a similar approach of targeting strong \ion{He}{2}-emitting young planetary nebulae in the southern hemisphere, future spectroscopic surveys are expected to uncover a significantly larger number of objects exhibiting Raman-scattered \ion{He}{2}.

Future observational strategies to study Raman-scattered features include spectropolarimetry and integral field unit ({\it IFU}) spectroscopy. \ion{He}{2} recombination lines are naturally expected to be completely unpolarized. However, if the expanding neutral shell adopts a bipolar nebular morphology, introducing non-spherical symmetry in the scattering region, strong linear polarization is anticipated for Raman-scattered \ion{He}{2}.
Due to the faintness of Raman-scattered \ion{He}{2}, conducting high-resolution spectropolarimetry is currently challenging. However, advancements in extremely large telescopes, such as the {\it Giant Magellan Telescope}, may enable such observations in the future \citep[e.g.,][]{ikeda04}.

The formation process of Raman-scattered \ion{He}{2} suggests that \ion{He}{2} emission is concentrated near the central hot star, whereas Raman-scattered \ion{He}{2} is prominent in the neutral region moving away from the star. This implies that spectra extracted from the central region of a PN may differ significantly from those obtained in the peripheral regions.  
By combining positional and velocity information, it may be possible to describe the various interactions between the ionized, neutral, and molecular components in these intriguing objects.

\begin{acknowledgments}
We are deeply grateful to the anonymous referee for his/her invaluable and constructive comments, which greatly enhanced the clarity and overall presentation of this paper. We thank the staff of the Bohyunsan Astronomical Observatory.
This work was also supported by the National Research Foundation of Korea (NRF) grants funded by the Korea government (No. NRF-2023R1A2C1006984). 
\end{acknowledgments}

\vspace{5mm}
\facilities{BOAO({\it BOES})}

\software{Cloudy \citep{2013RMxAA..49..137F} }
\bibliography{references}{}
\bibliographystyle{aasjournal}

\counterwithin{figure}{section}
\renewcommand\thefigure{\thesection\arabic{figure}}    
\appendix

\section{High Resolution Spectroscopy of Young PNe}\label{sec:appendix_observation}

In this appendix, we show our {\it BOES} spectra of young PNe belonging to Group~N and Group~H,
in which no Raman-scattered \ion{He}{2} was found.

\begin{figure}[h]
\centering
    \begin{subfigure}{}
    \includegraphics[scale=0.25]{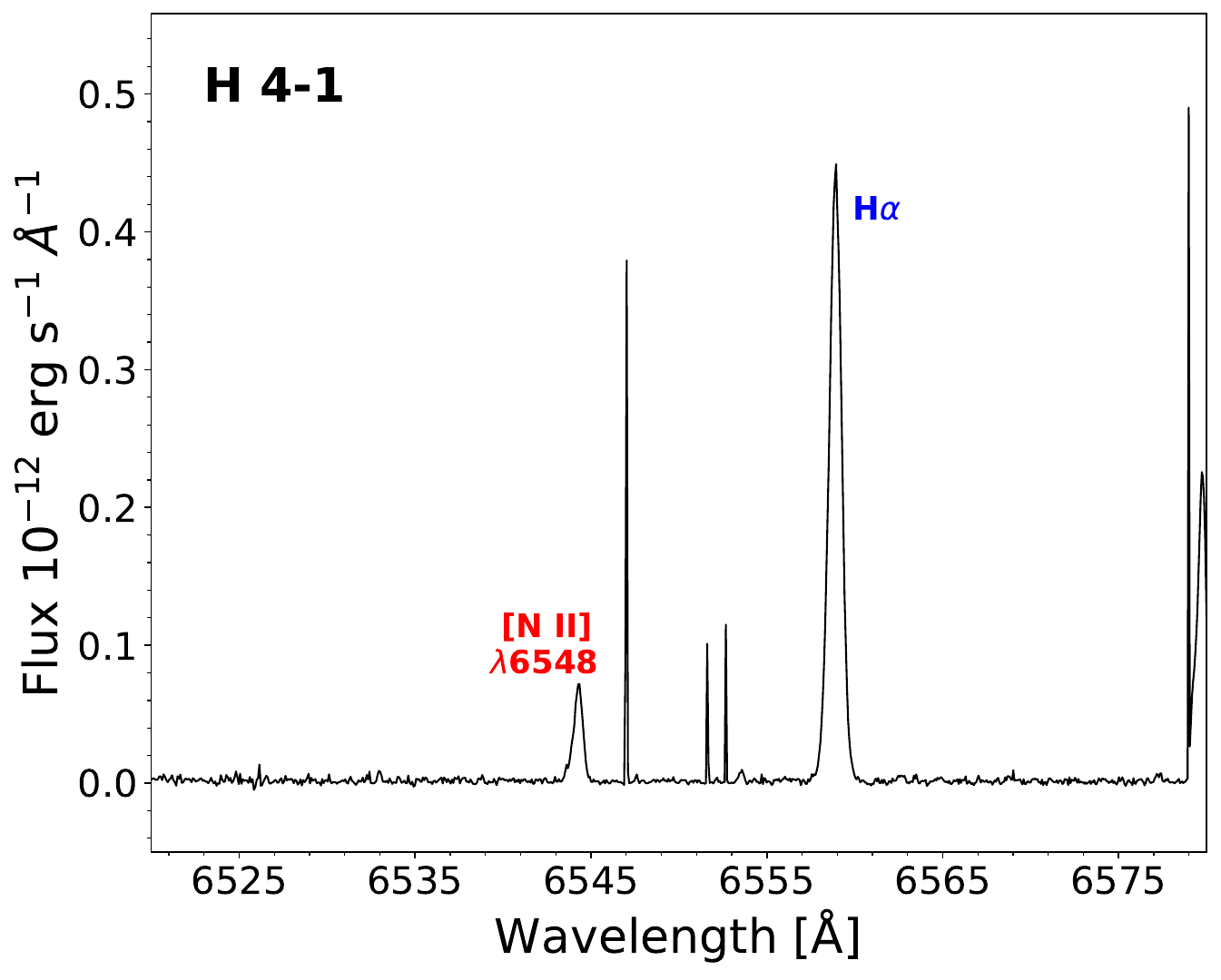}
    \end{subfigure}
    \begin{subfigure}{}
    \includegraphics[scale=0.25]{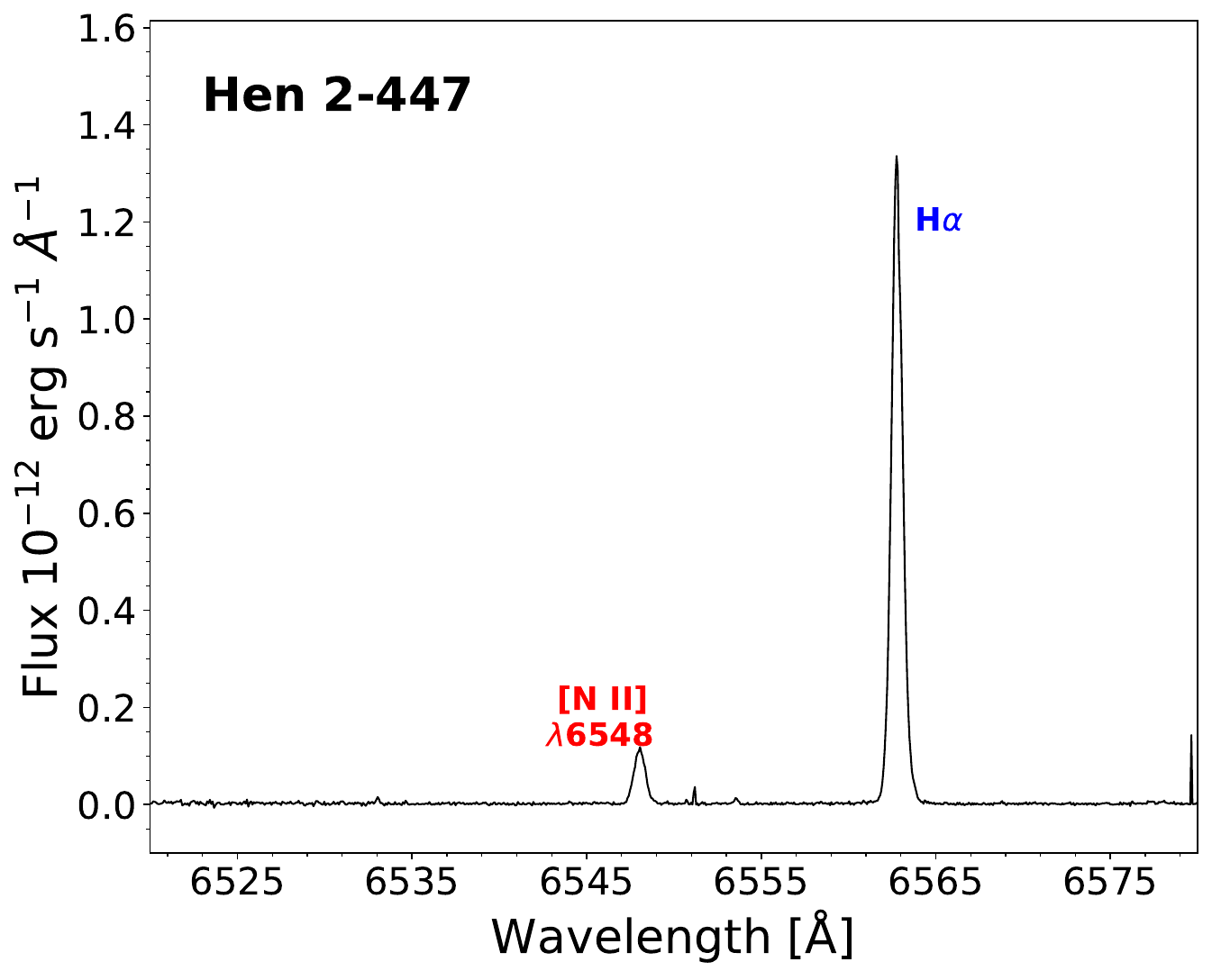}
    \end{subfigure}
    \begin{subfigure}{}
    \includegraphics[scale=0.25]{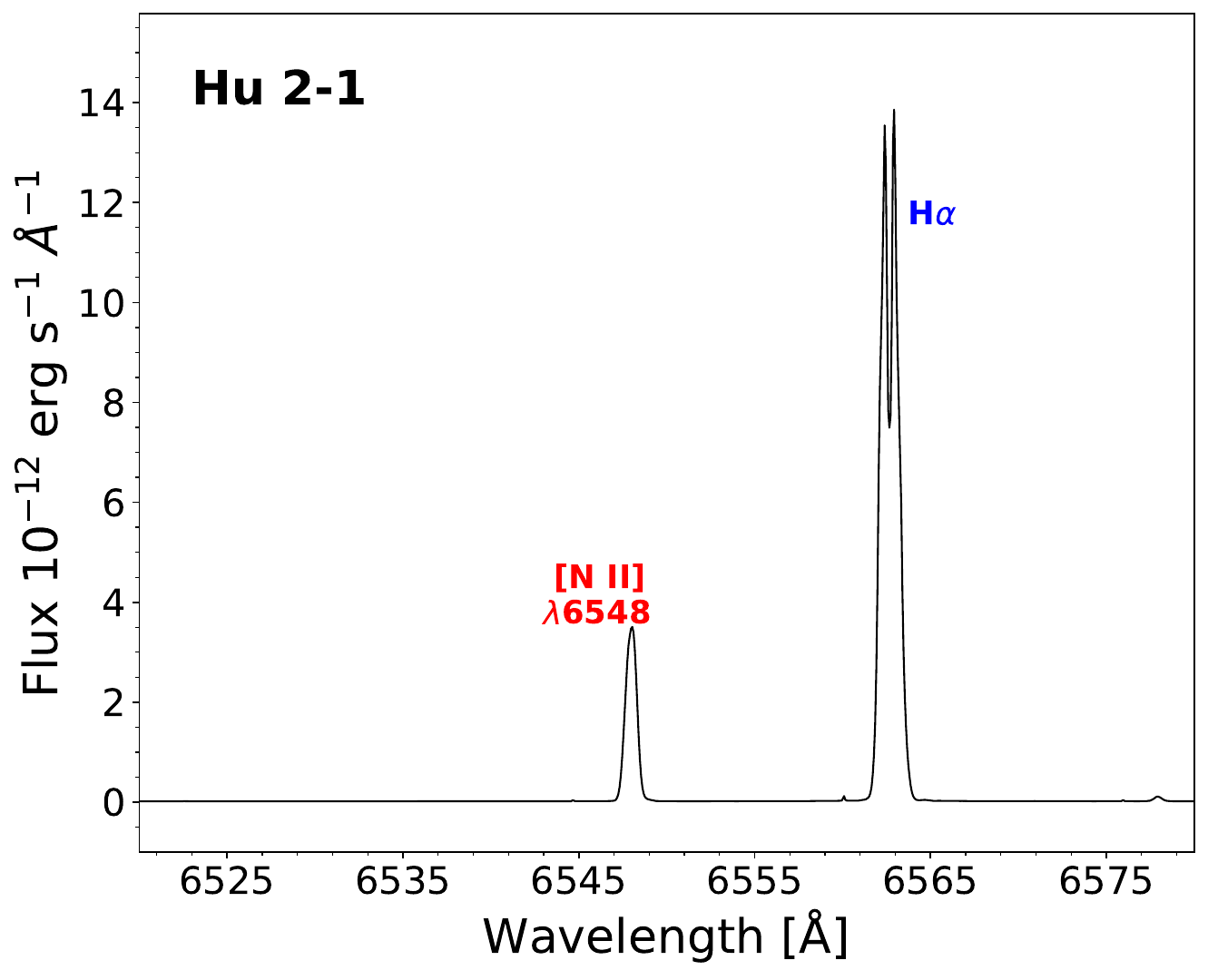}
    \end{subfigure}
\caption{{\it BOES} spectra around H$\alpha$ of the young PNe H~4-1, Hen~2-447, and Hu~2-1 in Group~N. Although these PNe are known to be \ion{He}{2} emitters in the literature, neither \ion{He}{2} $\lambda$6560 
nor \ion{He}{2} $\lambda$6527 is detected in our {\it BOES} data.}
\label{fig:group_n}
\end{figure}

In Figure~\ref{fig:group_n}, we show the {\it BOES} spectra of H~4-1, Hen~2-447, and Hu~2-1,
in which we find no \ion{He}{2} emission lines at 6560 \AA. Furthermore,
a close inspection of these objects shows that no \ion{He}{2}$\lambda$4686 is apparent in our {\it BOES} data.
For example, \cite{otsuka23} presented spectroscopy of H~4-1 using the Kyoto University Seimei 3.8 m telescope, in which a
clear presence of \ion{He}{2}$\lambda$4686 is noted. This strongly implies that our failure of \ion{He}{2}
detection in H~4-1 may be attributed to insufficient exposure in our observations.

\begin{figure*}
\centering
    \begin{subfigure}{}
    \includegraphics[scale=0.24]{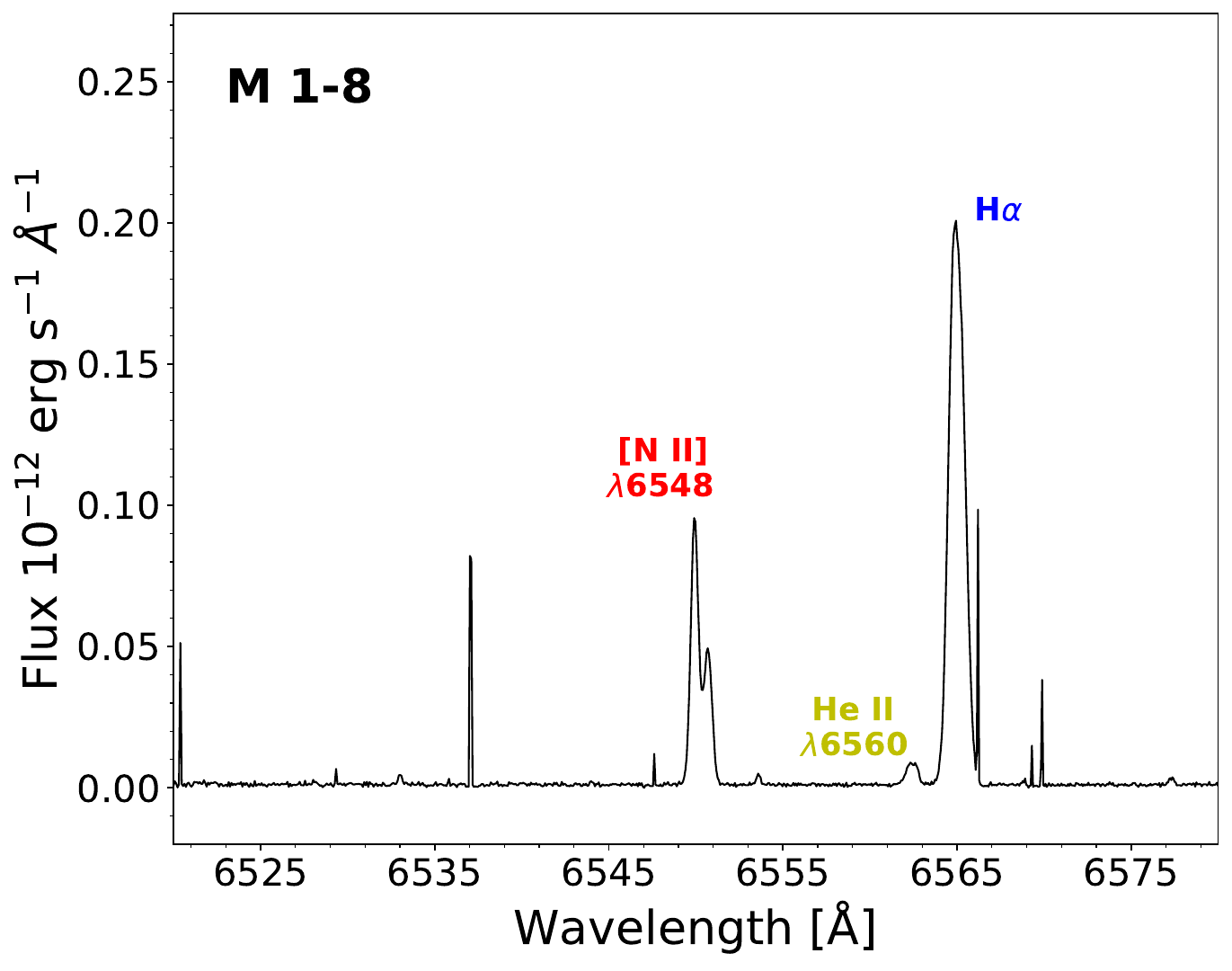}
    \end{subfigure}
    \begin{subfigure}{}
    \includegraphics[scale=0.24]{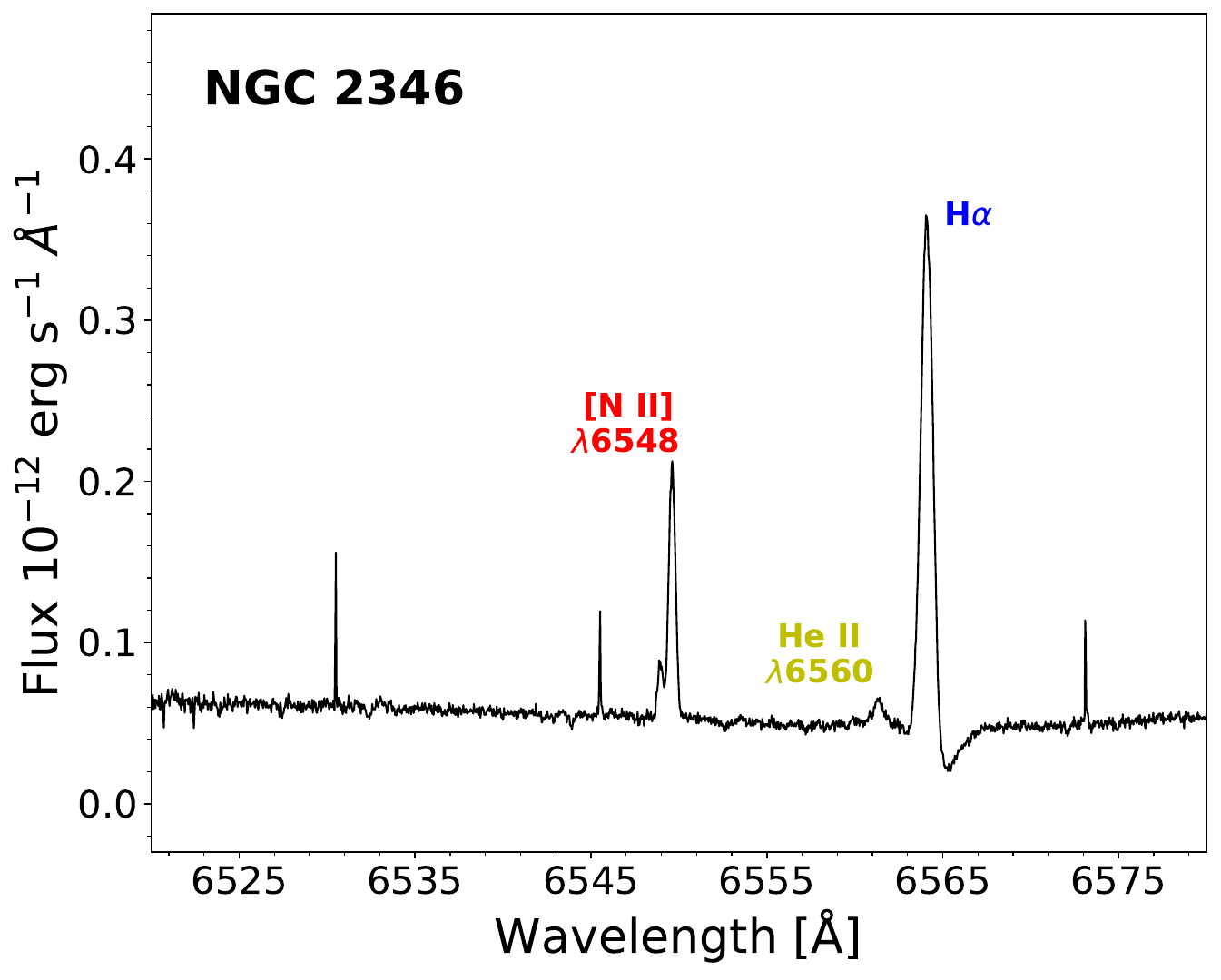}
    \end{subfigure}
    \begin{subfigure}{}
    \includegraphics[scale=0.24]{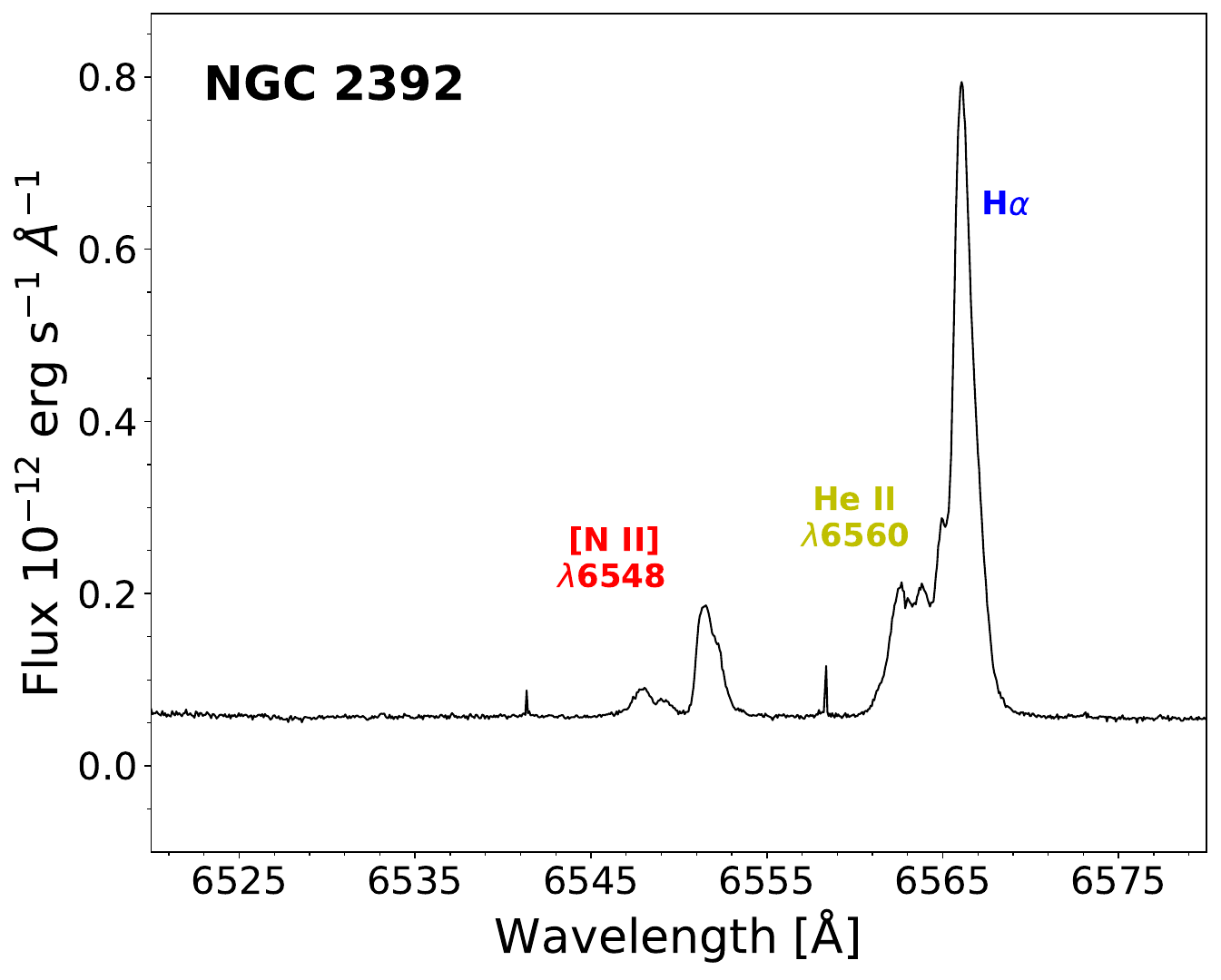}
    \end{subfigure}
    \begin{subfigure}{}
    \includegraphics[scale=0.24]{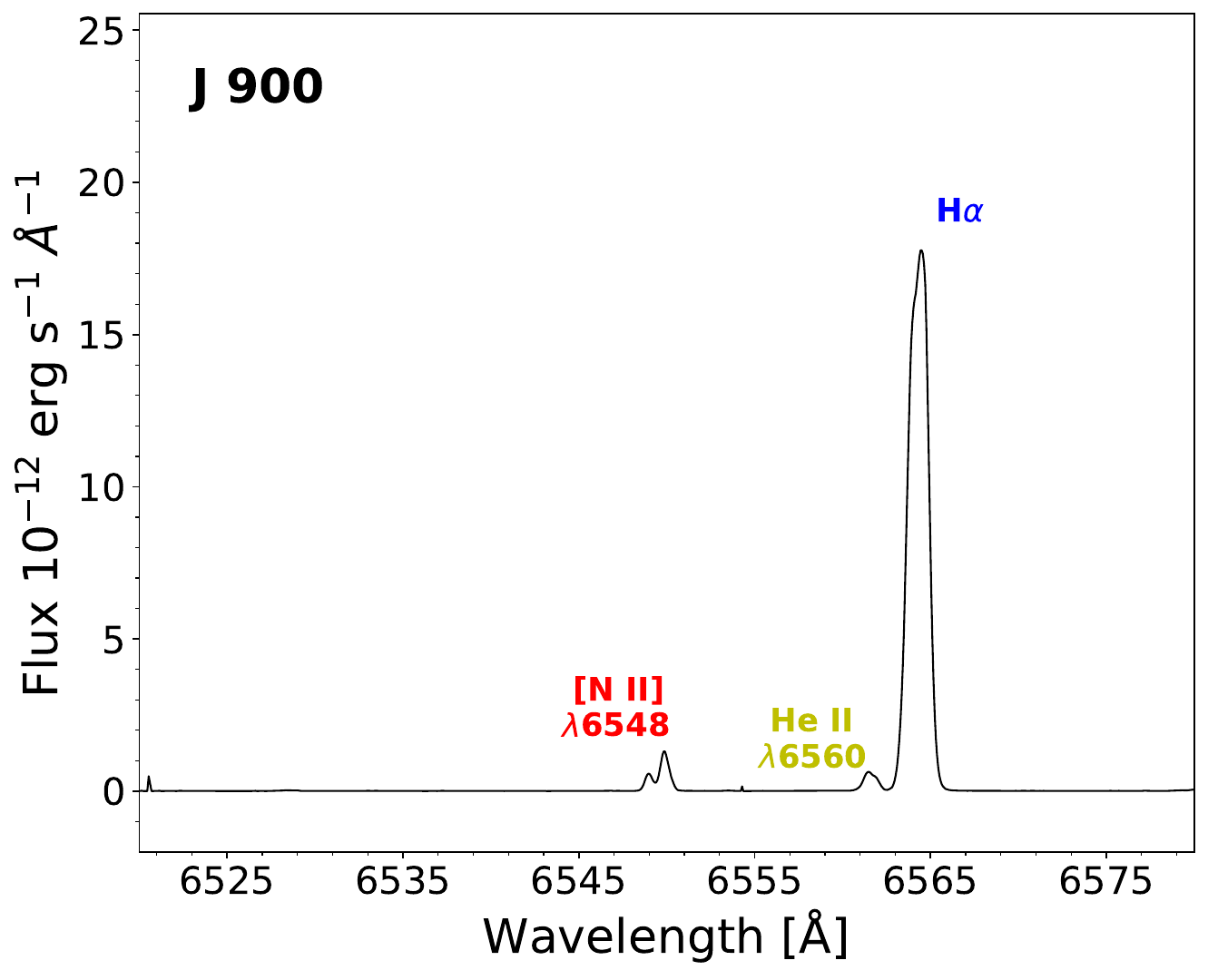}
    \end{subfigure}
        \begin{subfigure}{}
    \includegraphics[scale=0.24]{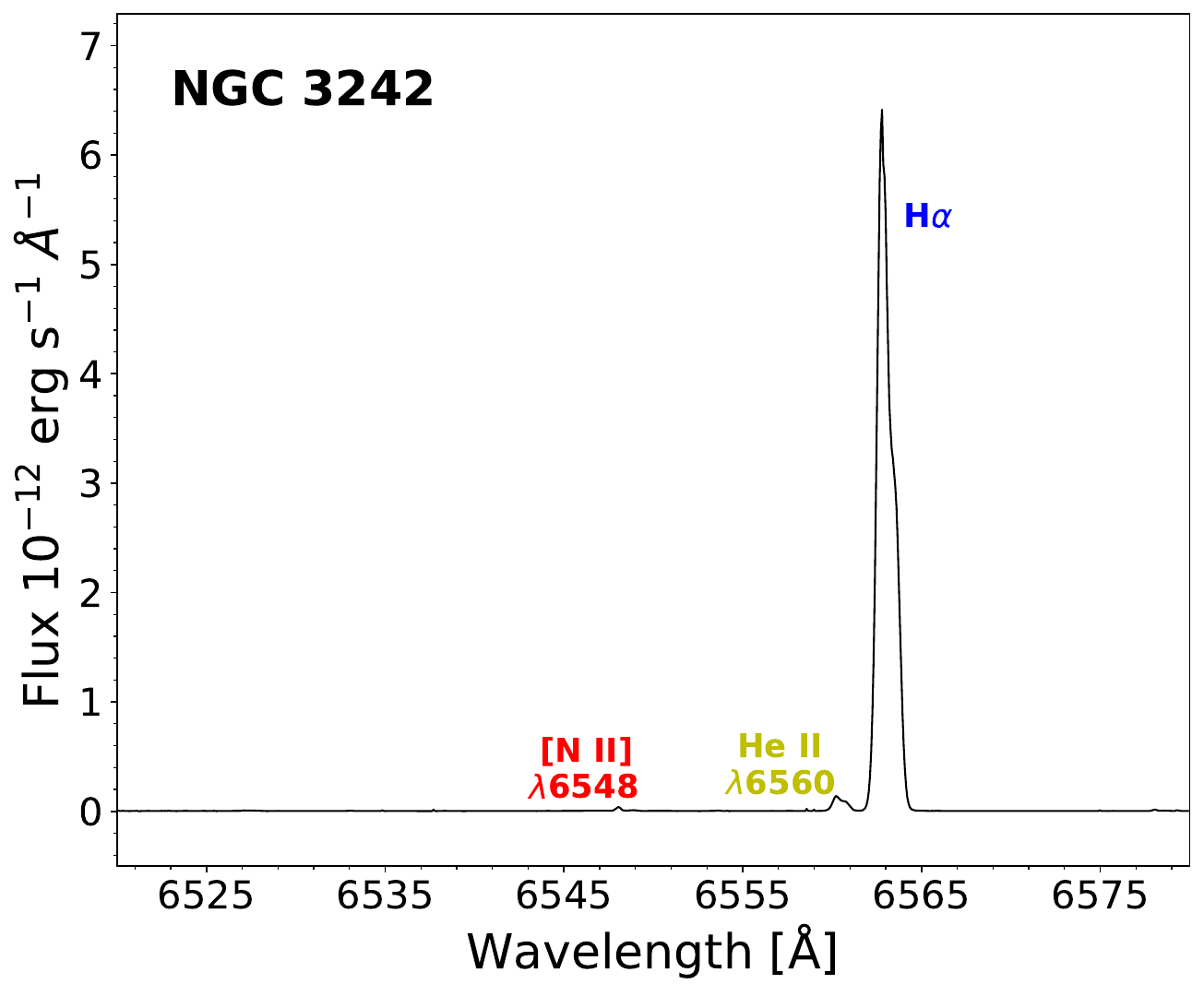}
    \end{subfigure}
\caption{{\it BOES} spectra of M 1-8, NGC~2346, and NGC~2392 (top panels), J~900, and NGC~3242 (bottom panels) in Group~H. In the top panels, \ion{He}{2} $\lambda$6560 line is visible but  \ion{He}{2} $\lambda$6527 is not discernible for these PNe. In the bottom panels, \ion{He}{2} $\lambda$6560 and \ion{He}{2} $\lambda$6527 lines are detected for these two PNe.}

\label{fig:group_w}
\end{figure*}

In the top panels of Figure~\ref{fig:group_w}, we show our {\it BOES} spectra of M~1-8, NGC~2346 and NGC~2392. In these
objects, we detect clearly \ion{He}{2}$\lambda$6560 emission line that appears as a blue shoulder of very strong
H$\alpha$. However, \ion{He}{2}$\lambda$6527 emission is not detected in these objects, implying insufficient
data quality to search for Raman-scattered \ion{He}{2}$\lambda$6545. It should also be
noted that the H$\alpha$ line of NGC~2392 is quite broad so that \ion{He}{2}$\lambda$6560 is embedded in the blue part of H$\alpha$. 

The lower two panels of Figure~\ref{fig:group_w} show {\it BOES} spectra of J~900 and NGC~3242, where both \ion{He}{2}$\lambda$6560 and
\ion{He}{2}$\lambda$6527 are clearly detected. In these objects, no wing features are noticeable around [\ion{N}{2}]$\lambda$6548, leading to nondetection of Raman-scattered \ion{He}{2}$\lambda 6545$.

\end{document}